\documentclass[prd,twocolumn,amsmath,amssymb,
footinbib,showpacs] {revtex4}
\usepackage{epsfig}
\usepackage{dcolumn}
\usepackage{bm}

\usepackage{amsmath}
\usepackage{amsfonts}
\usepackage{amssymb}

\newcommand {\nn} {\nonumber}
\newcommand {\half} {\frac{1}{2}}
\newcommand {\p} {\prime}

\newcommand{\bna}{\bar{\nabla}}

\newcommand{\wth}{\widetilde{h}}

\newcommand{\tnu}{\tilde{\nu}}
\newcommand{\tsigma}{\tilde{\sigma}}

\newcommand{\be}{\begin{equation}}
\newcommand{\ee}{\end{equation}}
\newcommand{\bea}{\begin{eqnarray}}
\newcommand{\eea}{\end{eqnarray}}
\newcommand{\JHEP}{J. High Energy Phys.}

\begin{document}

\author{Ishwaree P. Neupane\footnote{On leave from
Dept. of Physics, Tribhuvan Univ., Kathmandu}}


\title{\large \bf Thermodynamic and gravitational instability on
hyperbolic spaces}


\affiliation{ Department of Physics,
National Taiwan University,\\
Taipei 106, Taiwan, R.O.C.\\
{\scriptsize \bf ishwaree@phys.ntu.edu.tw, ~
Ishwaree.Neupane@cern.ch }\\
~~~~~}



\begin{abstract}

We study the properties of anti--de Sitter black holes with a
Gauss-Bonnet term for various horizon topologies ($k=0,\,\pm 1$)
and for various dimensions, with emphasis on the less well
understood $k=-1$ solution. We find that the zero temperature (and
zero energy density) extremal states are the local minima of the
energy for AdS black holes with hyperbolic event horizons. The
hyperbolic AdS black hole may be stable thermodynamically if the
background is defined by an extremal solution and the extremal
entropy is non-negative. We also investigate the gravitational
stability of AdS spacetimes of dimensions $D>4$ against linear
perturbations and find that the extremal states are still the
local minima of the energy. For a spherically symmetric AdS black
hole solution, the gravitational potential is positive and
bounded, with or without the Gauss-Bonnet type corrections, while,
when $k=-1$, a small Gauss-Bonnet coupling, namely,
$\alpha\ll{l}^2$ (where $l$ is the curvature radius of AdS space),
is found useful to keep the potential bounded from below, as
required for stability of the extremal background.

\end{abstract}


\maketitle

\section{Introduction}

In parallel with the development of AdS conformal field theory
(CFT) correspondence~\cite{Maldacena97a,Witten98}), black holes in
AdS space are known to play an important role in dual field
theory~\cite{Witten98a}.
It has also been learned that the Einstein equations when
supplemented by a negative cosmological constant admit black holes
as exact vacuum solutions, whose event horizons are hypersurfaces
${\cal M}$ with zero, positive or negative constant curvature
($k=0,\,+1$, or $-1$).
This ${\cal
M}$ may be related to the ${\cal M}^{\prime}$ on which the dual
field theory is defined by a rescaling of the metric.

Anti--de Sitter black holes with nonspherical event horizons have
been constructed in four and higher
dimensions~\cite{Vanzo,Birmingham98a,Mann96a}. Earlier work on
closely related AdS thermodynamics can be found
in~\cite{Lemos95a}. The analysis in~\cite{Birmingham98a} is well
motivated from the AdS/CFT correspondence.
Reference~\cite{Emparan99b} discusses AdS/CFT duals of $k=-1$
topological black holes in the spirit of a holographic counterterm
method developed in~\cite{Vijay99a,Emparan99a}. Here we study the
thermodynamic and gravitational stability of a class of
Gauss-Bonnet (GB) black holes in AdS space, which also have the
feature that the horizon (hypersurface) ${\cal M}$ is an
$(n-1)$-dimensional Einstein space with constant curvature
($k=-1,\,0,\,+1$).

There is the issue of the positivity of the total energy when AdS
black holes have non-spherical horizons. Usually, the positive
energy theorems show a stability of spacetimes which become
asymptotically either locally flat or anti--de Sitter. The known
theorems do not extend to the Horowitz-Myers
soliton~\cite{Horowitz98a}, whose AdS asymptotic is a toroidal
space with a zero constant curvature. The AdS soliton is a
nonsupersymmetric background but in the AdS/CFT context it is
conjectured to be a ground state for planar black holes; see also
Ref.~\cite{Galloway01a}.

Another interesting issue with hyperbolic AdS black hole
spacetimes is the choice of a background. In a gravitational
theory, it is necessary to make a Euclideanized action finite by
assigning classically stable lowest energy configurations to AdS
black hole spacetimes with a curvature $k=0,\,\pm 1$ of the
horizons. For $k=+1$, the background is simply a global AdS space,
which is the solution at finite
temperature~\cite{Hawking83a,Witten98a}. But, for $k=-1$, the
ground state may be different from a solution that is locally
isometric to a pure (global) AdS space~\cite{Vanzo,Birmingham98a}.

Hyperbolic AdS black holes are known to exhibit some new and
interesting features, such as an increase in the entropy that is
not accompanied by an increment in the
energy~\cite{Emparan99b,Emparan99a}. They are also relevant to
studying CFTs with less than maximal (or no)
supersymmetry~\cite{Emparan98a,Klemm99a}. In supergravity
theories, maximally symmetric hyperbolic spaces naturally arise as
the near-horizon region of certain $p$-branes~\cite{Kehagias00a}
and black hole geometries~\cite{Horowitz91a}. Thus the choice of a
ground state for the hyperbolic AdS black holes as well as their
thermodynamic and gravitational (or dynamical) stability are the
important issues.

For a spherically symmetric solution (i.e., $k=+1$), for instant,
the AdS Schwarzschild solution, the hypersurface ${\cal M}$ is
usually a round sphere, while, for hyperbolic AdS black holes,
${\cal M}$ is either a hyperbolic space $H^{n-1}$ or its quotient
$H^{n-1}/\Gamma$. Therefore, for $k=-1$, one reasonably assigns
the zero temperature (and zero energy) extremal state as a
background; see, for example, Refs.~\cite{Vanzo,Birmingham98a}. In
this paper, we further show, with or without a Gauss-Bonnet term,
that the extremal states are local minima of the energy for
hyperbolic AdS black holes.

In Ref.~\cite{Gibbons02a}, Gibbon and Hartnoll studied a classical
instability of spacetimes of dimensions $D>4$ against metric
perturbation, by considering generalized black hole metrics in
Einstein gravity. In this paper, we generalize those results by
including a Gauss-Bonnet term. We find that the black hole
spacetime whose AdS asymptotic is a hypersurface of negative
constant curvature could be unstable under metric perturbations if
the background is a zero mass topological black hole. We show
that, with or without a Gauss-Bonnet term, the extremal states are
local minima of the energy for $k=-1$ AdS spacetimes against
linear perturbations. We further argue that the extremal
background, defined with a negative extremal mass, can be
gravitationally (or dynamically) stable if the ground state metric
receives higher curvature corrections, like a GB term, with small
couplings.

The layout of the paper is as follows. In Sec. II we give black
hole solutions in AdS space, compute extremal parameters, and
define different reference backgrounds in Einstein gravity
modified with a Gauss-Bonnet term. In Sec. III we compute
Euclideanized actions applicable to the curvature $k=0,\,\pm 1$ of
the event horizons. In Sec. IV we relate the free energy and
specific heat curves and discuss the thermal phase transitions. In
Sec. V we turn to stability analysis of the background metrics
(vacuum solutions) in Einstein gravity under metric perturbations,
by setting up a Sturm-Liouville problem. We extend this analysis
in Sec. VI for the background metrics in Einstein-Gauss-Bonnet
theory. Section VII contains discussion and conclusion.

\section{Black holes in AdS Space}

Our starting point is the Lagrangian of gravity including a
Gauss-Bonnet term \bea\label{action1} I&=&\frac{1}{16\pi
G_{n+1}}\int
d^{n+1}x\,\sqrt{-g}\,\left(R-2\Lambda\right)\nn \\
&{}&+\,\alpha^\prime \int d^{n+1}x\,\sqrt{-g}\,
\left(R_{abcd}R^{abcd}-4 R_{cd}R^{cd}+R^2\right).\nonumber\\
\eea
Usually, the action is supplemented with surface terms (or a
Hawking-Gibbons type boundary action), which can be found, for
example, in Ref.~\cite{Myers87}, but they have no role in the AdS
black hole calculations~\cite{Hawking83a,Witten98a}.

The black hole solutions for the action~(\ref{action1}) were first
given by Boulware and Deser~\cite{Deser85a}, which were studied by
Myers and Simon~\cite{Myers88a}, within the context of Lovelock
gravity, by regularizing the classical action;
see~\cite{Wiltshire} for a discussion of charged Gauss-Bonnet
black holes. There has been considerable interest in generalizing
those solutions with $\Lambda < 0$ and $k \neq
1$~\cite{Cai01a,Nojiri01c,IPN02a,IPN02b}, and also within the
context of dimensionally extended Lovelock gravity~\cite{Cai98a}.

\subsection{Gauss-Bonnet black holes in AdS space}

The Einstein field equations modified by a Gauss-Bonnet term take
the following form \begin{equation} \label{newEE}
R_{ab}-\frac{1}{2}\,R g_{ab}+ \Lambda g_{ab}=16\pi
G_{n+1}\alpha^\p \left[\frac{1}{2}{\cal L}_{GB}
g_{ab}-2H_{ab}\right]\,, \end{equation} where $H_{ab}\equiv R
R_{ab}-2R_{acbd}R^{cd}+R_{acde}R_b\,^{cde}-2R_{ac} R_b^c$ and
${\cal L}_{GB}=R^2-4R_{ab}R^{ab}+R_{abcd}R^{abcd}$. For
$\alpha^\p>0$, we have the well known Gauss-Bonnet black hole
solution
\begin{equation}
ds^2=-\,f(r)\,dt^2+\frac{dr^2}{f(r)}+{r^2}\, d\Sigma_{k,n-1}^2\,,
\end{equation}
with
\begin{equation} \label{harmonic1}
 f(r)= k+\frac{r^2}{2\alpha}\pm \,
\frac{r^2}{2\alpha}\sqrt{1+\frac{8\alpha\Lambda}{n(n-1)}
+\frac{4\alpha\,\mu}{r^n}}\,, \end{equation} where $\alpha=16\pi
G_{n+1}\, (n-2)(n-3)\alpha^\prime$, and $\mu$ is an integration
constant. The metric of an $(n-1)$-dimensional space ${\cal M}$,
whose Ricci scalar equals $(n-1)(n-2)k$, is denoted as
$d\Sigma_{k,n-1}^2$; the latter is the unit metric on $S^{n-1}$,
$I\!\!R^{n-1}$, or $H^{n-1}$, respectively, for $k=1,\,0$, or
$-1$.

When $\alpha=0$, the cosmological constant is fixed as $\Lambda=-
n(n-1)/2{\ell^2}$, while, for $\alpha>0$, there is a rescaling,
namely, $\ell^2\to l^2= \ell^2/\left(1-\alpha/\ell^2\right)$. For
generality, henceforth, we use a common scale $l^2$, unless
otherwise stated, but the convention that $l^2\to \ell^2$ in the
limit $\alpha\to 0$ is to be understood. The dimension of $\alpha$
is $(\mbox{length})^2$.

\subsection{Extremal Gauss-Bonnet black holes}

For $\alpha>0$, the periodicity of the Euclidean time is
\begin{equation} \label{HawkingT} {\beta}=\frac{4\pi\, r_+\, l^2\,
\left(r_+^2+2\alpha\,k\right)} {n\,r_+^4+k\,(n-2)\,r_+^2\,l^2
+(n-4)\alpha\,k^2\,l^2}\,. \end{equation} When $k=-1$, $\beta$
starts from zero at the smallest radius $r_+=\sqrt{2\alpha}$,
except for the coupling $4\alpha=l^2$. The spacetime region can be
singular with no black hole interpretation if $4\alpha>l^2$ or/and
$r_+<\sqrt{2\alpha}$, and so one should be interested only in the
regions $4\alpha\leqslant l^2$ and $r_+^2>2\alpha$. The saturation
limit $4\alpha=l^2$ may be taken only if one also approaches the
critical limit $r_+^2=2\alpha$.

The parameter $\mu$ in Eq.~(\ref{harmonic1}) may be expressed in
terms of the horizon radius $r_+$, namely,
\begin{equation} \label{BHMass} \mu  =
r_+^{n-2}\left(k+\frac{r_+^2}{l^2} +\frac{\alpha
k^2}{r_+^2}\right)\equiv \frac{16\pi G_{n+1}\,M}{(n-1)V_{n-1}}\,,
\end{equation} where $M$ is the Arnowitt-Deser-Misner (ADM)
mass of a black hole, and $V_{n-1}$ is the volume of
$d\Sigma^2_{k,n-1}$. In the limit $\beta\to \infty$, the $k=-1$
extremal parameters are
\begin{eqnarray}
\mu_{extr}&=&\frac{2 r_{extr}^{n-2}}{n-4} \left(\frac{2}{n}-
\sqrt{\left(\frac{n-2}{n}\right)^2-\frac{4(n-4)\alpha}{n\,l^2}}\,
\right)\,,\label{mextr}\\
r_{extr}^2&=&\left(\frac{n-2}{2n}\right)l^2 \left(1+
\sqrt{1-\frac{4n(n-4)}{(n-2)^2}
\,\frac{\alpha}{l^2}}\right)\,.\label{rextr}
\end{eqnarray}
For $\alpha=l^2/4$, $\mu_{extr}=0$ for any $n$. It is somewhat of
a misnomer to call the $\mu_{extr}=0$ state an extremal state,
because extremal black holes are defined to have zero temperature,
which requires $\alpha<l^2/4$. Moreover, the solution with
$\mu_{extr}=0$ saturates the bound $r_+^2 \geqslant
2\alpha=l^2/2$, so the proper extremal black holes are those that
satisfy $\alpha < l^2/4$ and have zero Hawking temperature.

\subsection{Choice of backgrounds}

For $\alpha^\p=0$, the harmonic function $f(r)$ is defined by
\begin{equation} \label{bhmetric2}
f(r)=k+\frac{r^2}{l^2}-\frac{\mu}{r^{n-2}} \,.\end{equation} For
$k=0$, a zero mass ground state is still legitimate and is an
acceptable background~\cite{Vanzo,Birmingham98a}. However, within
the AdS/CFT context, Horowitz and Myers~\cite{Horowitz98a} have
proposed an AdS soliton as a candidate ground state for planar
black holes. The AdS soliton metric, for example, in five
dimensions, has the form
\begin{eqnarray}
ds^2&=&\frac{r^2}{l^2}\,\left(1-\frac{r_0^4}{r^4}\right)\,d\phi^2
+\frac{l^2}{r^2}\left(1-\frac{r_0^4}{r^4}\right)^{-1}\,dr^2 \nn
\\
&{}&+\, r^2
\left(-\,dt^2+\sum_{i=1}^{2}\left(d\theta^i\right)^2\right)\,,
\end{eqnarray}
where $r_0$ is a constant related to the AdS soliton mass. We are
interested here in the $k=-1$ case. When $\alpha=0$, $k=-1$, and
$n=4$, the extremal mass parameter is $\mu_{extr}=-\,l^2/4$, and
so the extremal metric has the form
\begin{eqnarray}
ds^2&=&-\, \frac{r^2}{l^2}\left(1-\frac{l^2}{2r^2}\right)^2
\,dt^2+\frac{l^2}{r^2}\left(1-\frac{l^2}{2r^2}\right)^{-2}\,
dr^2\nn \\
&{}& +\, r^2\,dH_{3,k=-1}^2\,.
\end{eqnarray}
For this to be a candidate ground state, the spacetime region
should be restricted to $r\geqslant r_e=l/\sqrt{2}$.

Next, consider the AdS black hole solution with $\alpha^\prime>0$:
\begin{equation} \label{newk=-1}
f(r)= k+\frac{r^2}{2\alpha}-
\frac{r^2}{2\alpha}\sqrt{1-\frac{4\alpha}{l^2}
+\frac{4\alpha\,\mu}{r^n}} \,. \end{equation} For $k=+1$, $\mu=0$
itself defines a reference background, and it is obvious that the
constraint $\alpha\leqslant l^2/4$ must hold. Interestingly, a
gravitational action defined with $\alpha=l^2/4$, in some
cases~\cite{Chamseddine89a}, is equivalent to the Einstein
gravity. In fact, the graviton propagators in {AdS}$_{n+1}$
spacetime, when $n\geqslant 4$, do not receive any corrections
from the massive (Kaluza-Klein) modes when $k=0$ and
$\alpha=l^2/4$ (see, for example, Ref.~\cite{IPN02e}), and so this
background may be stable under linear perturbations.

However, when $k=- 1$, stability of the background requires
$\alpha < l^2/4 $. That is, as in Einstein
gravity~\cite{Birmingham98a,Emparan99a}, the extremal mass
$\mu_{extr}$ takes only a negative value. In particular, when
$\alpha=l^2/4$ and $n=4$, the Ricci scalar and Kretschmann scalar
$K~(=R_{abcd}R^{abcd})$ read \begin{eqnarray}
R&=&-\,\frac{40}{l^2}+\frac{12\sqrt{\mu\,l^2}}{r^2 l^2}\,,\nn \\
K&=&\frac{160}{l^4}+\frac{48\mu}{r^4\,l^2}\mp
\frac{96\sqrt{\mu\,l^2}}{r^2\,l^4}\,.\end{eqnarray} The metric
spacetime is only asymptotically AdS when $\mu\neq 0$, and $\mu<0$
is not allowed in this case. Next, consider, for example, the
coupling $\alpha=l^2/12$. When $r$ is large, the curvature scalars
approximate to \begin{eqnarray} R&=&-\frac{40}{l^2}\,\Big(3\mp
\sqrt{6}\Big)-\frac{10(15\pm 4\sqrt{6})\,\mu}{r^4}+{\cal
O}\left(\frac{1}{r^6}\right),\nn \\
K&=&\frac{480(5\mp 2\sqrt{6})}{l^4}+{\cal
O}\left(\frac{1}{r^8}\right)\,,\end{eqnarray} in satisfying
$\mu>-\,2r^4/l^2$, $r^2>l^2/2\sqrt{3}$. $K$ diverges when
$\mu=-\,2r^4/l^2$, so $\mu_{extr}$ should be greater than this
value. The extremal solution is obtained after replacing $\mu$ in
Eq.~(\ref{newk=-1}) by $\mu_{extr}$. This is a physically
motivated choice of background for topological black
holes~\cite{Vanzo,Birmingham98a}.

\section{Background subtraction and thermodynamic
quantities}

For a solution of the equations of motion, the classical
action~(\ref{action1}) becomes
\begin{equation} \label{onshell}
I=\frac{1}{16\pi G_{n+1}}\int
d^{n+1}x\,\sqrt{-g}\left(-\frac{2}{n-3}\, R
+\frac{8\Lambda}{n-3}\right).
\end{equation} The Gauss-Bonnet term
is a topological for $n=3$, and so we must concern ourselves here
with spacetimes for which $n\geqslant 4$. The
action~(\ref{onshell}) diverges for a classical solution when
integrated from zero to infinity. To make the action (or energy)
finite, there should be a regularized setting -- a cutoff in the
radial integration and subtraction of a suitably chosen
background. For $k=+1$, the background is simply an AdS space with
$\mu=0$, while, for $k=-1$, the subtraction of a non zero mass
extremal background appears more
physical~\cite{Birmingham98a,IPN02b}.

For a pure anti--de Sitter space ($X_1$) any value of the
periodicity $\beta^\p$ is possible, while the black hole spacetime
($X_2$) has a fixed periodicity $\beta$. One may thus adjust
$\beta^\p$ such that the geometry of the hypersurface at $r=R\to
\infty$ is the same for AdS space and AdS-Schwarzschild space if
$k=+1$~\cite{Witten98a}, and for the extremal solution and
hyperbolic AdS black hole spacetime if
$k=-1$~\cite{Birmingham98a}. The surface terms have no role in the
large $r$ limit because the black hole corrections to the AdS
metric or extremal state vanish too rapidly at $r=R\to
\infty$~\cite{Hawking83a,Witten98a}. Thus, as in
Refs.~\cite{Witten98a,Birmingham98a}, we may fix $\beta^\p$ by
demanding
\begin{equation} \beta'\sqrt{k-\frac{\mu
_{extr}}{r^2}+\frac{r^2}{l^2}}\simeq \beta \sqrt{k-\frac{\mu
}{r^2}+\frac{r^2}{l^2}}\,. \end{equation} In doing this, we find
the (Euclidean) action difference $\widehat{I}=I(X_2)-I(X_1)$ to
be \begin{eqnarray} \widehat{I}&=&
-\frac{V_{n-1}\,r_+^{n-4}\,(n-1)\beta}{16\pi
G_{n+1}\,(n-3)} \Bigg[2\,k r_+^2-\mu \,r_+^{4-n}+\frac{4\,r_+^4}{l^2}
\nn \\
&{}&~~~ -\,\frac{8\pi r_+^3}{(n-1)\beta}\Bigg]
-\beta\,M_{extr}\delta_{k,-1}\,,
\end{eqnarray}
where $M_{extr}=(n-1)V_{n-1}\mu_{extr}/16\pi G_{n+1}$. [This can
be easily written, using Eq.~(\ref{BHMass}), in the form reported
earlier in Ref.~\cite{IPN02b}; cf., Eq.~(10)]. The free energy of
a black hole is given by $F=\widehat{I}/\beta$, namely,
\begin{eqnarray} \label{Free-energy}
F&=& \frac{(n-1) V_{n-1} r_+^{n-4}}{16\pi G_{n+1}(n-3)}
\left[\left(k {r_+}^2-\alpha {k}^2\right)
+\frac{3 {r_+}^4}{l^2}\right]\nn \\
&{}&+\,\frac{V_{n-1}r_+^{n-1}}{2(n-3)G_{n+1}}\frac{1}{\beta}
-M_{extr}\delta_{k,-1}\nonumber \\
&=&\frac{V_{n-1}\,r_+^{n-2}}{16\pi G_{n+1}}
\Bigg[\left(k-\frac{r_+^2}{l^2}\right) +\frac{n-1}{n-3}\,
\frac{\alpha k^2}{r_+^2}\nn \\
&{}&-\frac{2n}{n-3}\, \frac{\alpha k\left(2r_+^2+k
l^2\right)}{l^2\left(r_+^2+2\alpha
k\right)}\Bigg]-M_{extr}\delta_{k,-1} \,.\end{eqnarray} This is
modified from the result in Ref.~\cite{IPN02a} only by the last
term, which is non zero when $k=-1$ and $\alpha\neq l^2/4$. But,
since $M_{extr}$ is $r_+$ or $\beta$ independent, the black hole
entropy turned out to be the same as was given in~\cite{IPN02a}:
\begin{equation}
\label{GBentropy} {\cal S}=\beta^2\,\frac{\partial F}{\partial
\beta} =\frac{V_{n-1} r_+^{n-3}}{4G_{n+1}}
\left(r_+^2+\frac{2(n-1)\alpha\,k}{(n-3)}\right)\,.
\end{equation}
(See also Ref.~\cite{Cai01a} for an alternative derivation.) The
entropy flow is given by
\begin{eqnarray}
\label{entropyflow} d {\cal S}&=&\frac{(n-1)V_{n-1}
r_+^{n-4}}{4G_{n+1}} \Big(r_+^2+2\alpha\,k\Big)\nn\\
&=&\beta\,\frac{(n-1)V_{n-1}}{16\pi G_{n+1}} \Bigg[\frac{n\,
r_+^{n-1}}{l^2}+ (n-2)k r_+^{n-3}\nn \\
&{}&+\, (n-4)\alpha k^2 r_+^{n-5}\Bigg]\,.
\end{eqnarray}
In the second line above, we have used Eq.~(\ref{HawkingT}). Using
the thermodynamic relation $d{\cal S} =\beta\,dE$, and after
integrating with respect to $r_+$, we arrive at
\begin{equation}\label{totalE1} E=M+E_0\,. \end{equation} One
reads $M$ from Eq.~(\ref{BHMass}), and $E_0$ is an integration
constant. The energy $E$, obtained directly using the
Euclideanized action, after a lengthy but straightforward
calculation, takes a remarkably simple form
\begin{equation} \label{totalE2}
E=\frac{\partial\widehat{I}}{\partial \beta}=M
-M_{extr}\,\delta_{k,-1}\,.\end{equation} Comparing
Eqs.~(\ref{totalE1}) and (\ref{totalE2}), we easily identify that
$E_0=-\,M_{extr}\,\delta_{k,-1}\geqslant 0$. For $k=+1$, the
thermodynamic energy is given by the black hole mass $M$ and the
AdS solution with $\mu=0$ is the one with lowest action and
energy. This is usually not the case for $k=-1$, rather,
$E=-M_{extr}>0$ when $\mu=0$, $E=0$ when $\mu=\mu_{extr}$, and
$E>0$ otherwise. Importantly, the total energy is a positive,
concave function of the black hole's temperature for all values of
$k~(=0,\,\pm 1)$ when $M_{extr}<0$.


\subsection{Extremal state as the ground state}

As we mentioned in the Introduction, the hyperbolic {AdS}$_5$
black holes present some special features, such as that the total
(thermodynamic) energy
\begin{eqnarray}\label{5denergy2}
E&=&\frac{3V_3\,r_+^2}{16\pi
G}\,\left(-1+\frac{r_+^2}{l^2}+\frac{\alpha}{r_+^2}\right)
+\frac{3l^2\,V_3}{64\pi G}\left(1-\frac{4\alpha}{l^2}\right)\nn \\
&=&\frac{3\,V_3}{16\pi G}\left(-\, r_+^2+\frac{r_+^4}{l^2}
+\frac{l^2}{4}\right)\geqslant 0\, \end{eqnarray} is independent
of the coupling $\alpha$. This result may have some new and
interesting consequences in the field theory dual, if the latter
can exist with a GB term. We can perhaps use the relation
$$ \frac{\alpha^\p}{l^2}=\left(2 g_{YM}^2 N\right)^{-1/2}$$
(in units $16\pi G_5=1$, $\alpha^\p=\alpha/2$). Then a small
$\alpha^\p$ corresponds to the strong coupling limit (i.e.,
$g_{YM}^2 N$ is large.) The black hole entropy is given by
$$
{\cal S}=|\beta|\left(E-F\right).
$$
For $k=0$ or $+1$, we find $\mbox{lim}_{\beta\to \infty}\beta F=
0$. This is usually not the case for a hyperbolic {AdS}$_5$ black
hole, since the free energy and the entropy both depend on the
coupling $\alpha$ but the energy $E$ (or energy density $E/V_3$)
does not. So, in particular, for a small (or vanishing) coupling
$\alpha^\p$, we find $\mbox{lim}_{\beta\to \infty}\beta F=
\mbox{finite}$. This result obviously extends to the related
discussions made in Ref.~\cite{Emparan99b}; more generally, the
variation of free energy (and entropy) when going from strong to
weak coupling limits.

\begin{figure}[ht]
\begin{center}
\epsfig{figure=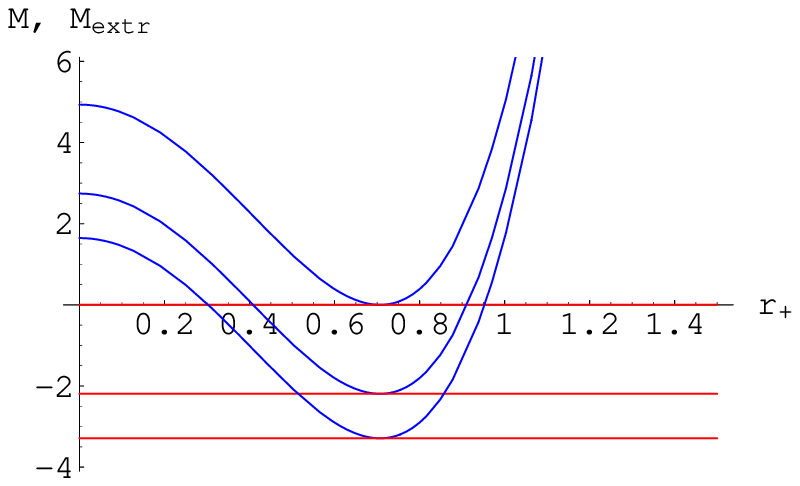, height=3.5cm, width=6.5cm}
\end{center}
\begin{center}
\epsfig{figure=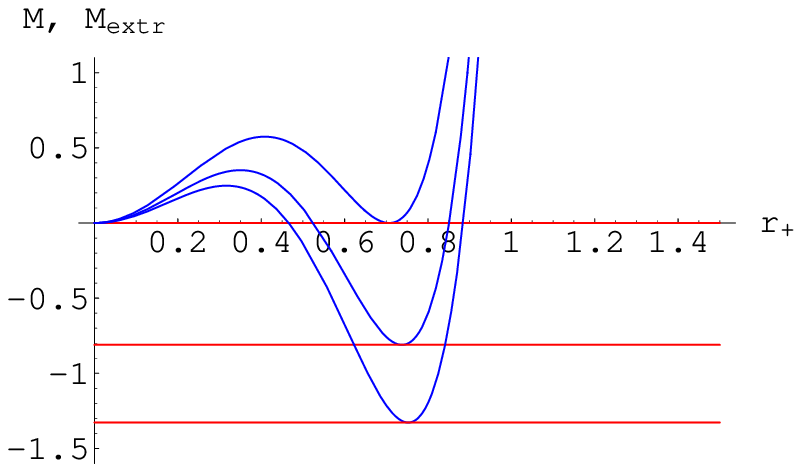, height=3.5cm,width=6.5cm}
\end{center}
\caption{The ADM mass $M$ (curved lines) and extremal mass
$M_{extr}$ (horizontal lines) as functions of the horizon. The
values are fixed at $l=1$, $16\pi G=1$, and (upper plot) $n=4$,
$V_3=2\pi^2$, $k=-1$, and $\alpha=1/4,~1/12,~0$ (top to bottom);
(lower plot) $n=6$, $V_3=\pi^3$, $k=-1$, and
$\alpha=1/4,\,17/100,\,0$ (top to bottom).} \label{figure1}
\end{figure}

\begin{figure}[ht]
\begin{center}
\epsfig{figure=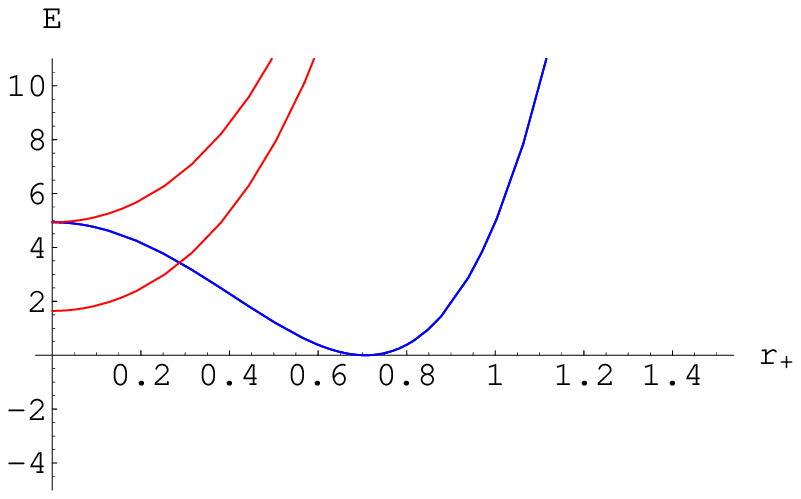, height=3.5cm, width=6.5cm}
\end{center}
\begin{center}
\epsfig{figure=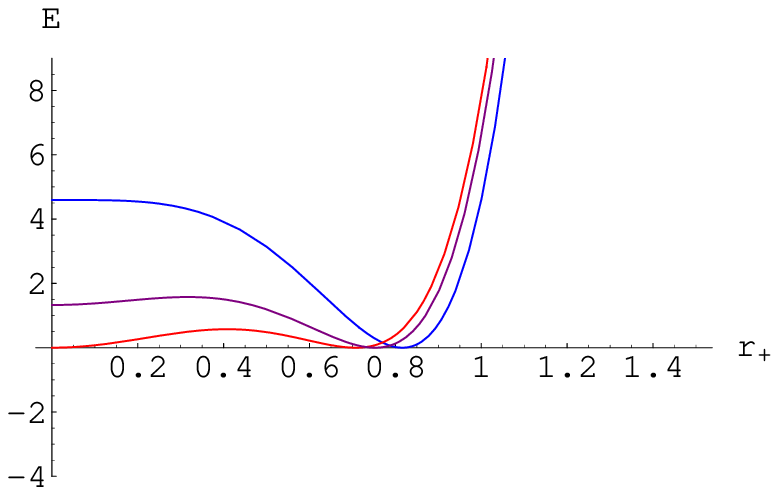, height=3.5cm,width=6.5cm}
\end{center}
\caption{The energy $E$ as a function of horizon ($r_+$). The
values are fixed at $l=1$, $16\pi G=1$, and $n=4$ (upper plot),
$V_3=2\pi^2$, $k=+1$ (curves that grow with $r_+$) at $\alpha=1/4,
1/12$ (upper to lower), and $k=-1$ (curve that has minimum at
$r_+=r_{extr}$); $n=6$ (lower plot), $V_3=\pi^3$, $k=-1$, and
$\alpha=1/4, 17/100, 0$ (top to bottom along the $E$-axis).}
\label{figure2}
\end{figure}

The total energy of hyperbolic black holes depends on the coupling
$\alpha$ when $n>4$, and it has a local minimum at the extremal
horizon position, which is seen also from the plots in
Figs.~\ref{figure1} and \ref{figure2}. Even for the coupling
$\alpha=l^2/4$, the total energy, for example, when $n=6$,
\begin{equation} E=\frac{5V_{5}}{16\pi G_{7}} \left( -\,
r_+^4+\frac{r_+^6}{l^2} +\frac{ r_+^2 l^2}{4}\right)
\end{equation}
is vanishing at the extremal horizon $r_+=l/\sqrt{2}$. But, since
$E$ is negative in the range $0<r_+<l/\sqrt{2}$, a massless
extremal state can violate the positive energy theorem, but a
negative mass extremal state with $\alpha<l^2/4$ always respects
the energy condition $E\geqslant 0$.

\subsection{Thermodynamic instability for $k=-1$}

Here we adopt a Euclidean path integral formulation along with a
consideration that the (extremal) entropy and specific heat are
non-negative at the background. One also notes that the black hole
entropy is always positive for the $k=\{0,1\}$ family. However,
for $k=-1$, the inequality $r_+^2\geqslant 2\alpha$, which must
hold in order to have a black hole interpretation, does not
guarantee that the (extremal) entropy and specific heat are always
positive, rather one must satisfy $r_+^2\geqslant
2(n-1)\alpha/(n-3)$. There is indeed an upper bound in the
coupling constant $\alpha$ which ensures a thermodynamical
stability; e.g., when $n+1=7$, the positivity of extremal entropy
requires $r_+^2>10\,\alpha/3$ and hence $\alpha/l^2<17/100$. In
particular, when one approaches the $\mu_{extr}=0$ limit at
$4\alpha=l^2$, so $M_{extr}=0$, the specific heat could be
negative, which mimics thermodynamic instability of the solution.

In the canonical ensemble, which should be the case here as we are
considering uncharged black holes, the second derivative of the
Euclidean action $\widehat{I}~(=\beta E-{\cal S})$ along the path
parameterized by $x$ is \begin{equation}
\left(\frac{\partial^2\hat{I}}{\partial^2\,
x}\right)_T=\frac{1}{\beta^2}\,\left(\frac{d\beta}{d x
}\right)^{2}\, \frac{d E}{d T}\,, \end{equation} where $x=x(T)$ is
the parameter that labels the path in the Euclidean path integral
formulation. Thus, as discussed in~\cite{Reall01a}, the black hole
is not the local minimum of the action when the specific heat
$dE/dT$ is negative.

\section{Specific heat and free energy curves and thermal
phase transition}

In Einstein gravity ($\alpha^\p=0$), small spherical black holes
have negative specific heat but large size black holes have
positive specific heat~\cite{Gregory93a}. There exists a
discontinuity of the specific heat as a function of temperature at
$r_+=l/\sqrt{2}$, and so small and large black holes are somewhat
disjoint objects. However, this is not essentially the case when
$\alpha^\p$ is nontrivial, and specially, in the spherical
{AdS}$_5$ case, the small black holes also have positive specific
heat~\cite{IPN02a} (see also the discussion in
Ref.~\cite{Cai01a}).

Here we study the {AdS}$_5$ and {AdS}$_7$ black holes, which are
of some particular interest in string or M theory as they may
provide duals for CFTs describing the world-volume theory for $N$
parallel $D3$- or $M5$-branes.

\subsection{Specific heat for hyperbolic black holes}

\begin{figure}[ht]
\begin{center}
\epsfig{figure=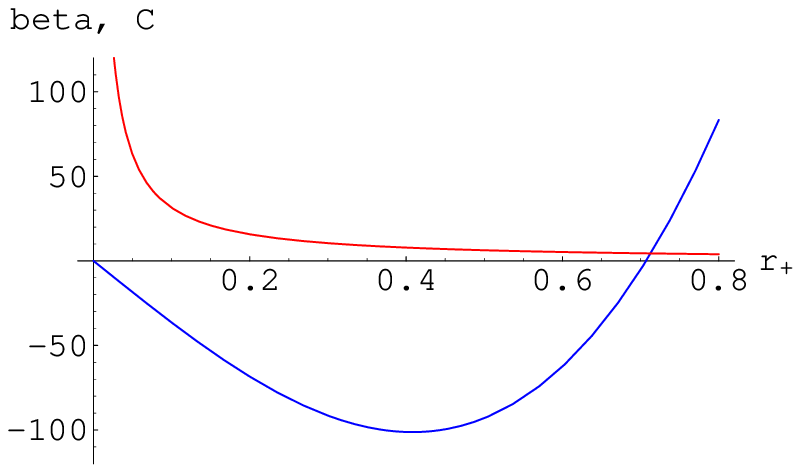,height=3.5cm,width=6.5cm}
\end{center}
\begin{center}
\epsfig{figure=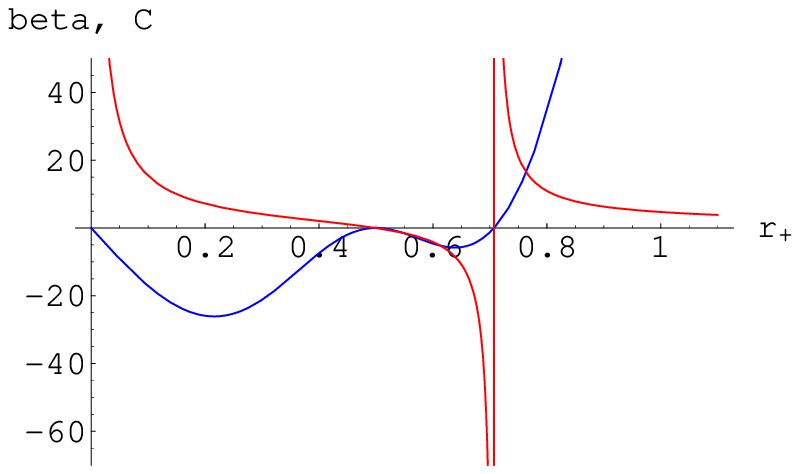,height=3.5cm,width=6.5cm}
\end{center}
\begin{center}
\epsfig{figure=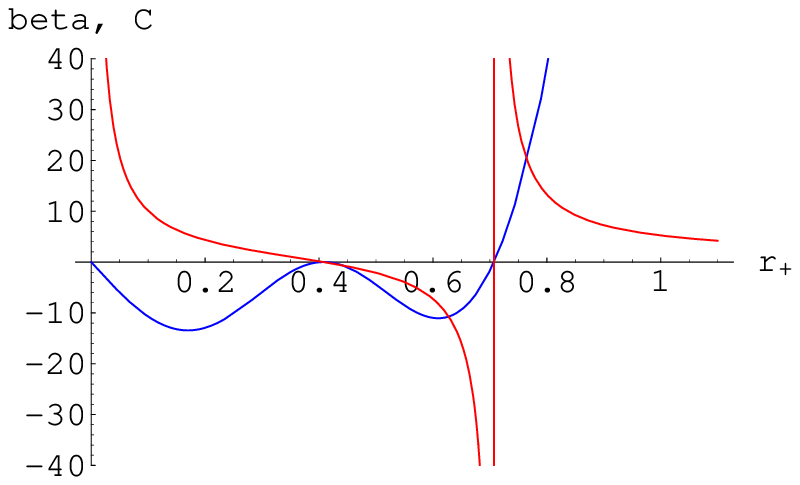,height=4.0cm,width=6.5cm}
\end{center}
\caption{The Euclidean period $\beta$ (curves that asymptote to
the $C$ or/and $r_+$ axes) and specific heat (curves with one or
more cusps) {vs} horizon position $r_+$. The values are fixed at
$l=1$, $n=4$, $V_3/16\pi G=2\pi^2$, $k=-1$, and $\alpha=1/4$,
$\alpha=1/8$ and $\alpha=1/12$ (top to bottom plots).}
\label{figure3}
\end{figure}

In the {AdS}$_5$ case, the specific heat curve has a single branch
for $\alpha=l^2/4$ and two branches for $\alpha < l^2/4 $. The
first cusp on the left, which will almost coincide with the $r_+$
axis for $\alpha\ll{l}^2$, has negative specific heat, so is
unstable. Moreover, the entropy formula
\begin{equation} \label{entropy5} {\cal
S}=\frac{V_{3,k=-1}\,r_+^{3}}{4G_{5}}
\left(1-\frac{6\alpha}{r_+^2}\right)\, \end{equation} shows that
the extremal entropy is non-negative only if $\alpha\leqslant
l^2/12$. It is then relevant to ask what would happen in the limit
$l^2/4>\alpha>l^2/12$. As a specific example, for $\alpha=l^2/8$,
the period $\beta=\pi
l^2\left(4r_+^2-l^2\right)/[2r_+\left(2r_+^2-l^2\right)]$ is
negative only in the range $0.5<r_+<0.707$. But when
$\alpha=l^2/8$, the entropy in Eq.~(\ref{entropy5}) becomes
negative in the range $l^2/2<r_+^2<3l^2/4$, although $\beta$ is
positive there. Thus, hyperbolic {AdS}$_5$ black holes are
thermodynamically stable only if $\alpha\leqslant l^2/12$, i.e.,
when ${\cal S}_{extr}\geqslant 0$ (see Fig.~\ref{figure3}).

\begin{figure}[ht]
\begin{center}
\epsfig{figure=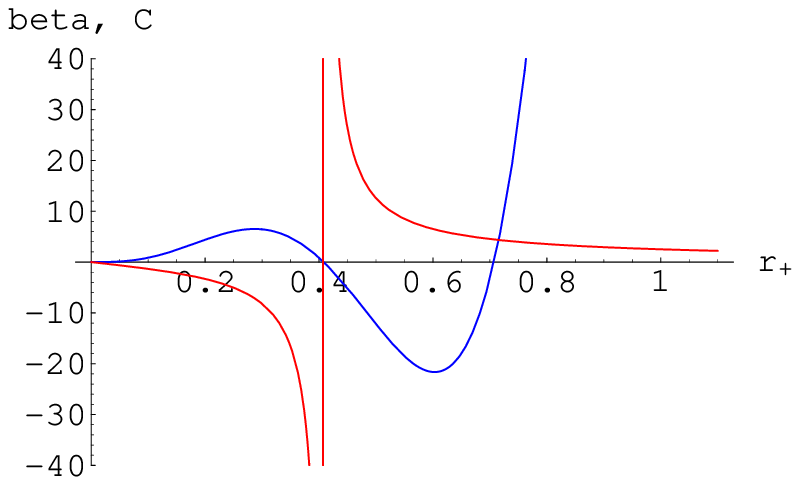,height=3.5cm,width=6.5cm}
\end{center}
\begin{center}
\epsfig{figure=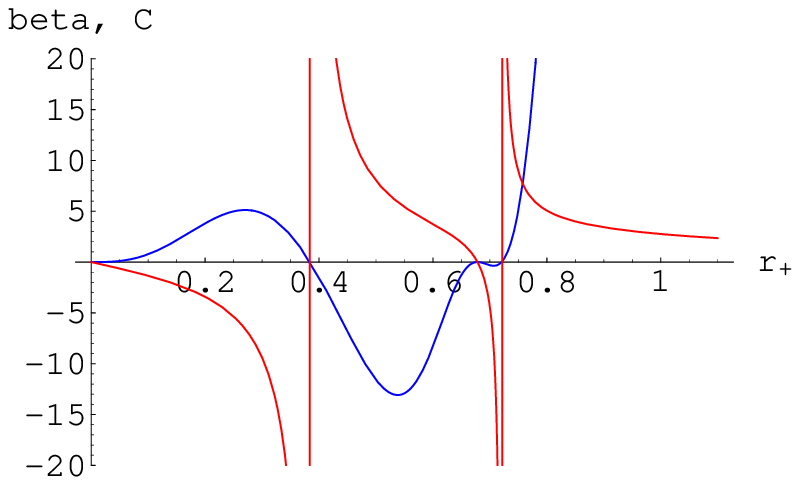,height=3.5cm,width=6.5cm}
\end{center}
\begin{center}
\epsfig{figure=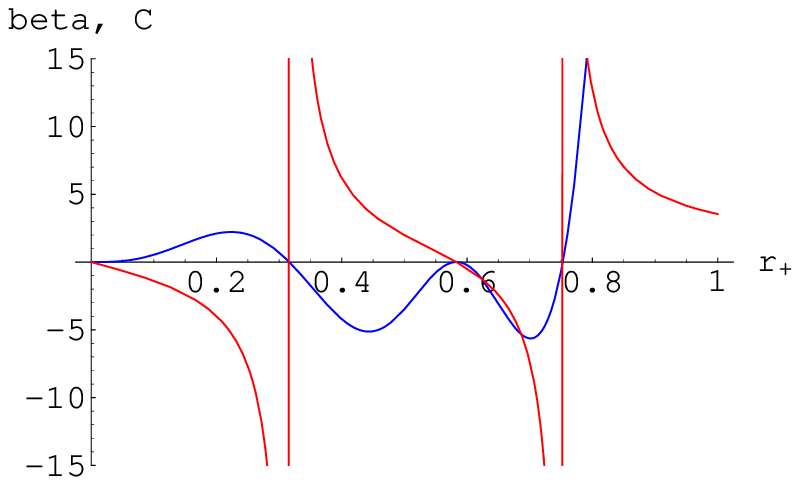,height=3.5cm,width=6.5cm}
\end{center}
\caption{The Euclidean period (curves that asymptote to the $C$
or/and $r_+$ axes) and specific heat (curves with two or more
cusps) {vs} horizon position $r_+$. The values are fixed at $l=1$,
$n=6$, $V_5/16\pi G=\pi^3$, $k=-1$, and $\alpha=0.25$,
$\alpha=0.23$, and $\alpha=17/100$ (top to bottom plots).}
\label{figure4}
\end{figure}

The free energy of a black hole is background dependent, while the
Hawking temperature is not. But the formulas $C=dE/dT$ and
$C=dM/dT$ give the same answer since $M_{extr}$ is $r_+$ or $T$
independent. In the {AdS}$_5$ case, the specific heat can be
negative for a small $r_+$, specially, if $\alpha>l^2/12$. This
particular feature is, although not totally new, different from
the $k=+1$ case. For $k=+1$, a thermodynamic instability may arise
due to a finite size effect, viz., a black hole of size
$r_+\leqslant l/\sqrt{2}$ is unstable, $r_+>l$ is stable, and
$l>r_+>l/\sqrt{2}$ is only locally preferred but globally
unstable. However, for $k=-1$, instability may arise from the both
a large coupling effect ($\alpha\sim l^2$) and a small size effect
($r_+<l$).

For the {AdS}$_7$ case, the extremal entropy is zero when
$\alpha=17 l^2/100$, for which the Euclidean period is
\begin{equation} \beta =\frac{4\pi r_+ l^2(50
r_+^2-17l^2)}{(10r_+^2-l^2)(30r_+^2-17 l^2)}\,.\end{equation} This
is negative when $0<r_+<0.3162$ and $0.5831<r_+<0.7528$. The free
energy is also negative in the latter range. That is, the first
and third cusps in the bottom plot of Fig.~\ref{figure4} should be
mirror reflected. There is no black hole interpretation for the
first cusp. For $\alpha=l^2/4$, there are only two cusps, because
in this case two unstable branches (the second and third cusps
that appear for $\alpha<l^2/4$) merge to a single cusp, which has
negative specific heat. We easily see that the thermodynamic
stability of hyperbolic black holes in $n=6$ requires
$\alpha/l^2\leqslant 17/100$. As a result, the specific heat and
extremal entropy are positive at the background.

Moreover, for $k=-1$, the entropy of a black hole may appear to be
negative, typically in the strong coupling limit $\alpha\sim l^2$;
see, for example,~\cite{Nojiri02a}. But this parameter space must
be excluded as an unphysical region because such a state is not
stable classically nor do the supergravity approximations allow
one to take $\alpha$ in the same order as $l^2$. As a result, for
$\alpha\ll{l}^2$, the hyperbolic black hole is still stable and
has positive entropy.

\subsection{Free energy for hyperbolic black holes}

In Ref.~\cite{Klemm99a}, it was argued that the stringy
corrections of order ${\alpha^\p}^3{R}^4$ do not give to rise a
thermal phase transition for flat and hyperbolic horizons, as a
function of temperature. But using an AdS soliton as the thermal
background of AdS black holes with Ricci flat horizons, a phase
transition that is dependent not only on the Hawking temperature
but also on the black hole area was found in~\cite{Surya01a}. This
may imply that for the flat space $I\!\!R^{n-1}$ compactified down
to a torus, and possibly also for the quotient of hyperbolic space
$H^{n-1}/\Gamma$, the choice of ground state is crucial to see a
possible phase transition. Here we want to use the obtained
thermodynamic quantities to determine a thermal phase structure in
the Einstein-Gauss-Bonnet gravity.

\begin{figure}[ht]
\begin{center}
\epsfig{figure=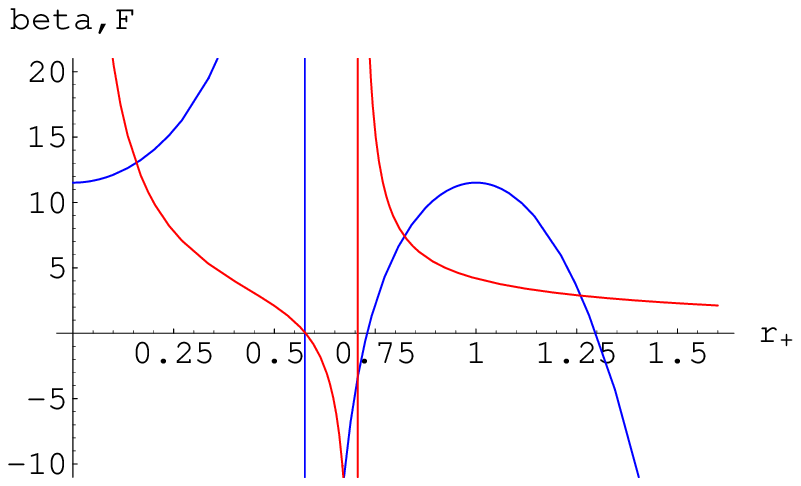, height=4.0cm, width=6.5cm}
\end{center}
\begin{center}
\epsfig{figure=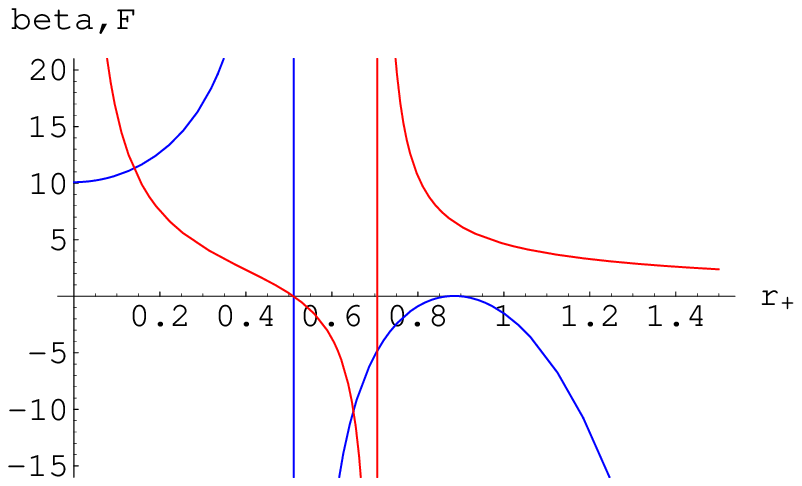, height=3.5cm,width=6.5cm}
\end{center}
\begin{center}
\epsfig{figure=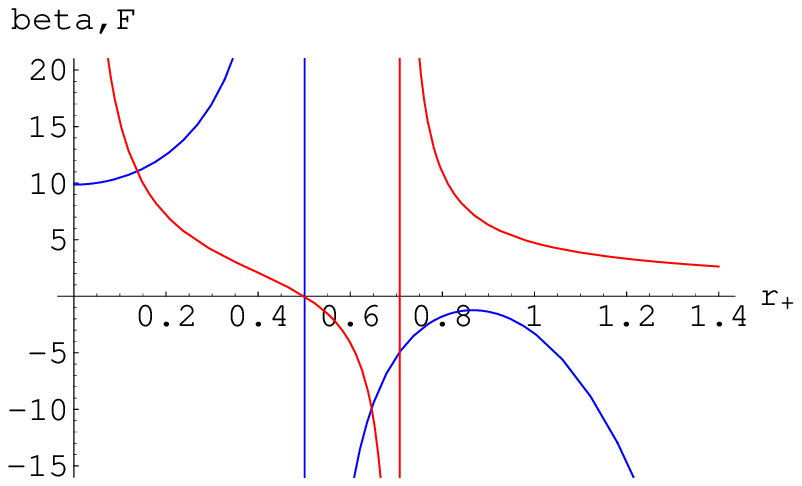,height=3.5cm,width=6.5cm}
\end{center}
\caption{The Euclidean period (curves that asymptote to the $F$ or
$r_+$ axis) and free energy (curves that are bounded from above or
take finite values at $r_+=0$) {vs} horizon position $r_+$. The
values are fixed at $k=-1$, $n=4$, $l=1$, $16\pi G=1$,
$V_3=2\pi^2$, and, $\alpha=1/6,\,0.1305,\,1/8$ (top to bottom
plots).} \label{figure5}
\end{figure}

In the limit $0<r_+^2\leqslant 2(n-1)\alpha/(n-3)$, the behavior
of extremal entropy is somewhat exotic, and so the second law of
thermodynamics may not hold. In addition, for a hyperbolic
{AdS}$_5$ black hole, the free energy scales with the coupling
$\alpha$, but the energy (or energy density) does not
(Fig.~\ref{figure5} and Eq.~(\ref{5denergy2})). The total energy
$E$ is vanishing at the extremal horizon, where $\beta= \infty$,
but the free energy can be positive, zero, or negative there
depending upon the coupling $\alpha$. As a result, the entropy is
non zero at the extremal state, in particular, for a zero or small
coupling $\alpha$. This result is consistent with the earlier
observations made in~\cite{Emparan99a,Emparan99b} with
$\alpha^\p=0$; a common thread in these results is that
$\mbox{lim}_{\beta\to \infty} \beta F=$ finite. The meaning of a
non zero extremal entropy may be traced back to the observation of
``precursor'' states first made in~\cite{Susskind99a}, when
studying a short distance behavior of AdS$\slash$CFT duality.

\begin{figure}[ht]
\begin{center}
\epsfig{figure=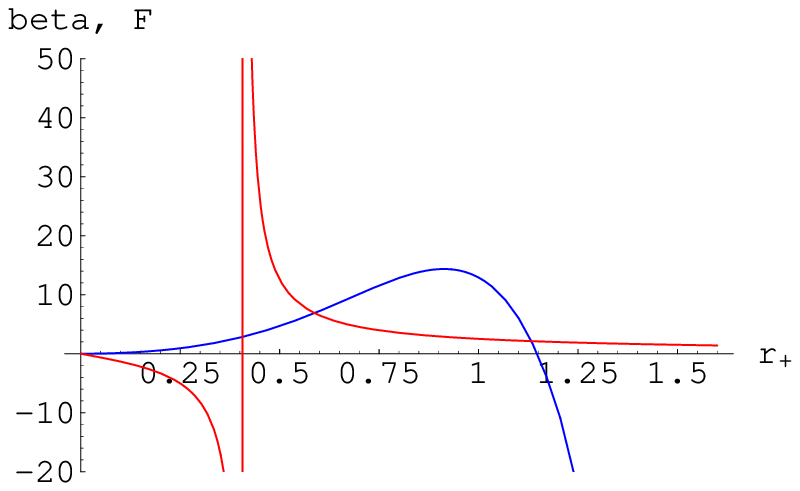, height=3.5cm, width=6.5cm}
\end{center}
\begin{center}
\epsfig{figure=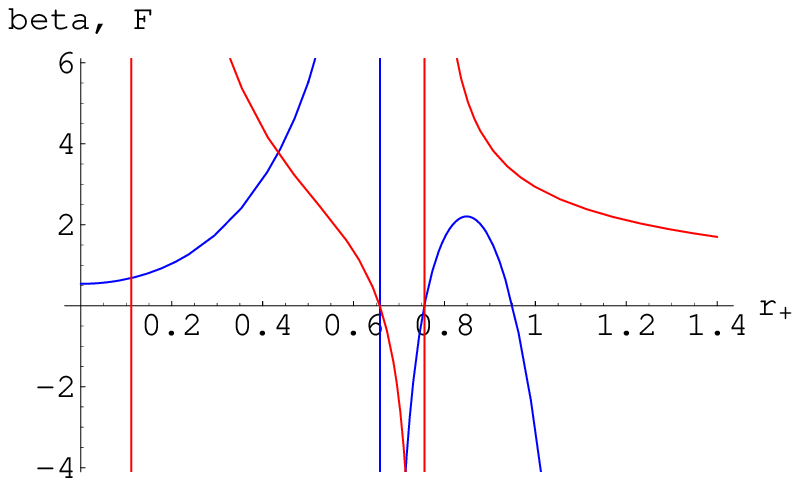, height=3.5cm,width=6.5cm}
\end{center}
\begin{center}
\epsfig{figure=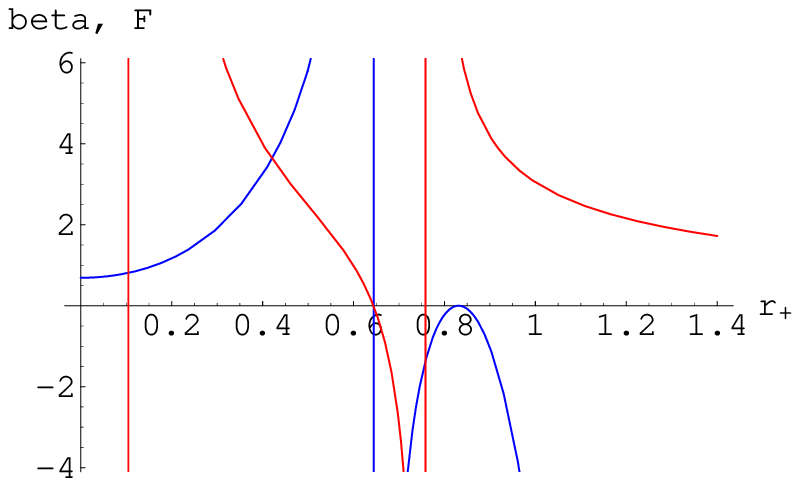, height=3.5cm,width=6.5cm}
\end{center}
\caption{The Euclidean period (curves that asymptote to the $F$
or/and $r_+$ axis) and free energy (curves that have finite values
at $r_+=0$ or bounded from above) {vs} horizon. The values are
fixed at $k=-1$, $n=6$, $l=1$, $16\pi G=1$, $V_5=\pi^3$, and
$\alpha=0.25, 0.2162, 0.2072$ (top to bottom plots).}
\label{figure6}
\end{figure}

The first plot in Fig.~\ref{figure6} corresponds to the
$\mu_{extr}=0$ state with $\alpha=l^2/4$. In the lower two plots,
the extremal state is shown by an asymptotic on the right, and the
critical state, which is absent in AdS$_5$ (cf.
Fig.~\ref{figure5}), is shown by another asymptotic on the left.
The middle vertical line, where the free energy diverges and
$\beta=0$, corresponds to the horizon at $r_+^2=2\alpha$. For
$\alpha<l^2/4$, the singularity at $r_+=\sqrt{2\alpha}$, where
$\beta=0$, is hidden inside the extremal horizon, so is harmless.

When $n=4$ and $0.25\,l^2>\alpha>0.1305\,l^2$, the free energy
appears to be positive in a certain range $r_2>r_+>r_{extr}$,
where the Hawking-Page temperature is finite, but, for a small
coupling $\alpha\ll{l}^2$, the free energy takes only the negative
value for positive $\beta$. It takes a maximum value at the
extremal state if ${\cal S}_{extr}=0$ (for example, at
$\alpha=l^2/12$ when $n=4$). There is no thermal AdS phase for the
coupling $\alpha<0.1305\,l^2$, and hence no phase transition
appears to occur in this case.

\subsection{Free energy for spherical black holes}

The spherical Gauss-Bonnet black holes were studied
in~\cite{Cai01a,IPN02a} in some detail, so we will be brief here.
For this case, since $E=M$, the free energy is vanishing when
\begin{equation} \label{stableF}
\alpha=\tilde{\alpha}= r_+^2
\left[\frac{4\sqrt{n(n-3)}-(n-3)}{6(n-1)}\right].
\end{equation}
This separates two other $F=0$ trajectories~\cite{Cai01a}:
\begin{eqnarray} \alpha_{\pm}&=&\pm \frac{r_+^2}{2}
\sqrt{\frac{9\,r_+^4}{l^4}+\frac{(15-n)\,r_+^2}{(n-1)\,l^2}
-\frac{(n+1)(7n-25)}{4(n-1)^2}}\nn \\
&{}&
+\,r_+^2\left(\frac{3\,r_+^2}{2\,l^2}-\frac{(n-7)}{4(n-1)}\right)
\,. \end{eqnarray}
Obviously, $n\geqslant 4$ is needed to ensure that $\alpha>0$. For
example, when $n=4$, we find $r_+^2=18\alpha/7>2\alpha$. The
$\alpha_-$ branch of the solutions gives a stable AdS solution
when $\alpha\ll{l}^2$, while the $\alpha_+$ branch may be
unstable.

\begin{figure}[ht]
\begin{center}
\epsfig{figure=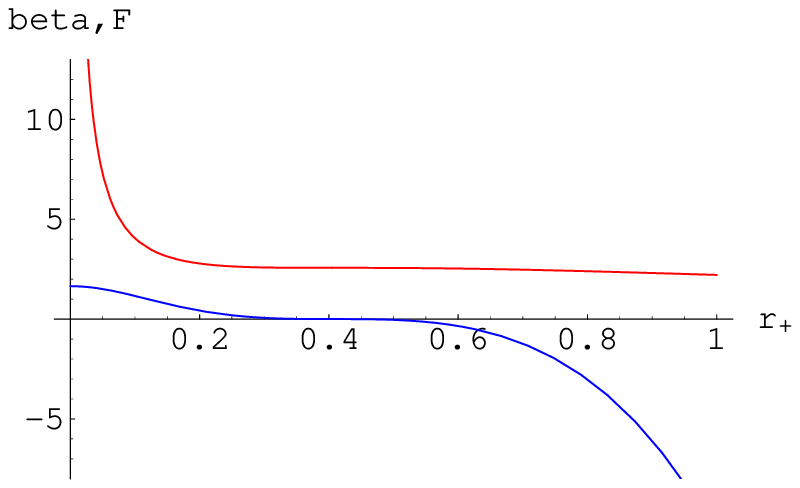,height=3.5cm,width=6.5cm}
\end{center}
\begin{center}
\epsfig{figure=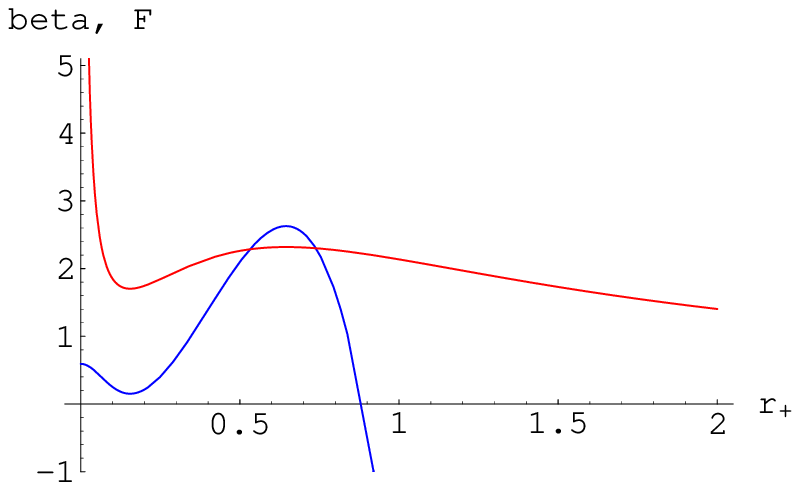,height=3.5cm,width=6.5cm}
\end{center}
\caption{Inverse temperature $\beta$ (curves that asymptote to the
$F$ and $r_+$ axes) and free energy (curves that cross $r_+$ axis)
{vs} horizon position $r_+$ for spherical black holes. The values
are fixed at $k=+1$, $n=4$, $l=1$, and, from top to bottom,
$\alpha=0.0278$, $0.01$. A new branch of a stable black hole
appears in the AdS$_5$ case.} \label{figure7}
\end{figure}
\begin{figure}[ht]
\begin{center}
\epsfig{figure=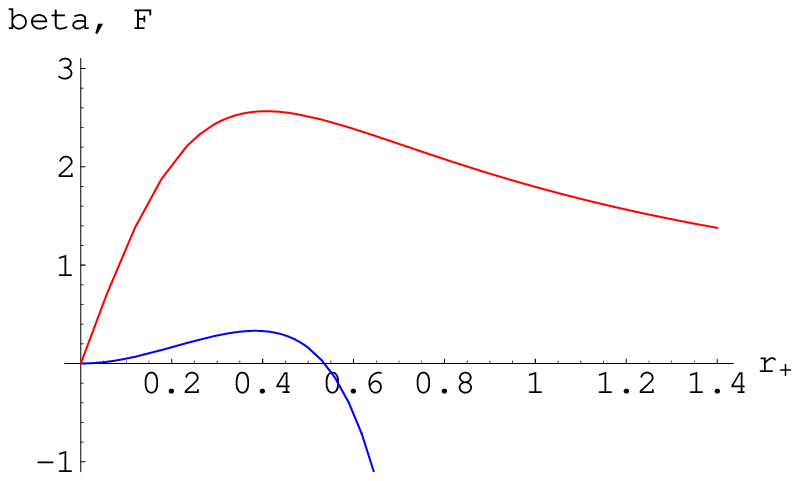,height=3.5cm,width=6.5cm}
\end{center}
\begin{center}
\epsfig{figure=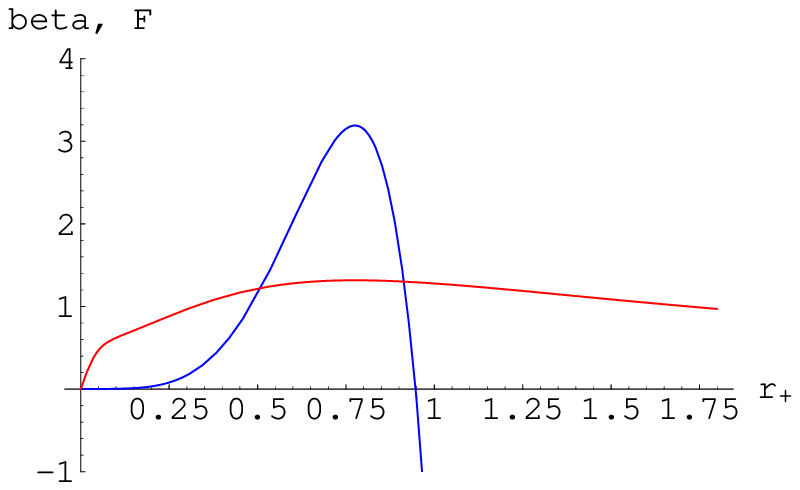,height=3.5cm,width=6.5cm}
\end{center}
\caption{Inverse temperature $\beta$ (curves that asymptote to the
$r_+$ axis) and free energy (curves that cross the $r_+$-axis)
{vs} horizon position $r_+$ for spherical black holes. The values
are fixed at $k=+1$, $n=6$, $l=1$, and, from top to bottom,
$\alpha=0.1$, $\alpha=0.01$.} \label{figure8}
\end{figure}

As in the $\alpha=0$ case, a spherical Gauss-Bonnet black hole
presents an interesting thermal feature, namely, the Hawking-Page
phase transition at finite temperature. In the {AdS}$_5$ case, as
$\alpha$ is decreased, the free energy lowers toward zero at low
temperature. For $\alpha\ll{l}^2$, in the small $r_+$ region, $F$
nearly approaches zero but it never touches the $F=0$ axis, e.g.,
when $\alpha/l^2=3\times 10^{-7}$, $F\sim 4\times 10^{-6}$. That
is, the free energy curve crosses the $F=0$ axis only once,
namely, when $r_+\lesssim l$, which corresponds to the
Hawking-Page phase transition point.

For $\alpha/l^2\gtrsim 0.0278$, the free energy is a monotonically
decreasing function of $r_+$ and the second branch (peak) on the
right disappears. Using the relation $ {\alpha^\p}/l^2=1/\sqrt{2
g^2 N}$,
we see that the Hawking-Page
transition disappears at sufficiently small 't Hooft coupling
$g_{}^2 N$.

As for the charged AdS black holes~\cite{Chamblin99a}, the inverse
temperature and free energy curves in Fig.~\ref{figure7} remind us
of the behavior of a van der Waals gas in the Clapeyron $(V,P)$
phase, where the point of inflection in $\beta(r)$ signals a
critical point. This behavior may be seen for spherical AdS black
holes that receive stringy corrections of the form ${\alpha^\p}^3
R^4$~\cite{Klemm99a} or $R_{abcd}R^{abcd}$~\cite{IPN02c}.

One also notes that, in all dimensions $n> 4$, the behavior of a
Hawking-Page transition is qualitatively similar for small and
large couplings, except that the free energy becomes more positive
as $\alpha$ is decreased. But still, as in the {AdS}$_5$ case, a
low temperature phase corresponds to a thermal AdS space with
$F\geqslant 0$ and a high temperature phase corresponds to an AdS
black hole with $F<0$.

\section{Stability analysis: metric perturbations}

The stability of background metrics under gravitational
perturbations can normally be checked from an analysis of whether
bound states with a negative eigenstate exist. If the
corresponding Schr\"odinger equations allow a state with $E<0$
and/or there are growing (quasi) normal modes at the event
horizon, and/or the potential is negative and unbounded from
below, then the spacetime metrics can be unstable.

\subsection{Linearized field equations}

Consider a linear perturbation of the metric
$$
\bar{g}_{ab}\to g_{ab}= \bar{g}_{ab}+h_{ab}\,,\quad
g^{ab}=\bar{g}^{ab}-h^{ab} $$  with $\vert h^a\,_b\vert \ll{1}\,.$
For $\alpha^\p=0$, the linearized equations are
\begin{equation} \delta R_{ab}=-\,n c\, h_{ab}\,, \end{equation}
where $c$ has the dimension of $\ell^2$. For $\alpha^\p>0$ and
$k=+1$, as evaluated in the Appendix, the perturbations $\delta
g_{ab}~(=h_{ab})$ must satisfy
\begin{eqnarray}
&&\left(1-2 C\alpha\right) \delta R_{ab}= -\,n\,C\,
\left(1-\frac{(3n^2-3n+2) C
\alpha}{n(n-3)}\right)h_{ab}\nn \\
&&~~~~~~~-\,\frac{2C\alpha}{n-3}\,\bar{g}_{ab}\left[\bna^c\bna^d
h_{cd}-\bna^2 h+C(n+1)h\right].\nonumber\\ \label{A3}
\end{eqnarray} with
\begin{equation}
\label{valueC} C=\frac{1}{2\alpha}\left(1\mp \sqrt{1+
\frac{8\alpha\Lambda}{n(n-1)}}\,\right)\equiv
\frac{1}{2\alpha}\left(1\mp \sqrt{1- \frac{4\alpha}{l^2}}\,\right)
.
\end{equation}
For transverse trace-free (de Donder) perturbations, $h_c^c=\bna^a
h_{ab}=0$, we obtain \begin{eqnarray} 2\delta
R_{ab}&=&\left(\Delta_L h\right)_{ab}\nn \\
&=&-\,\frac{2n C}{1-2C\alpha}\left[1-\frac{C \alpha
(3n^2-3n+2)}{n(n-3)}\right]h_{ab} \,.\nn \\
&{}& ~~~~ \end{eqnarray} If the base ${\cal M}$ is hyperbolic,
then it is not {\it a priori} sufficient to prove stability for
the tensor modes only. In this case, the scalar and vector modes
can also be unstable but we will not consider such modes here.

We wish to study the stability of the background metrics under the
conditions
\begin{equation} h_{0a}=h_{1a}=0\,.
\end{equation} With these constraints, along with a requirement
that the base ${\cal M}$ is itself an Einstein space
$\widetilde{R}_{i j}(h)=(n-2)k\,\tilde{g}_{i j}$, the transverse
trace-free property of $h_{ab}$ is characterized by $h_{i j}$,
where $i,\,j,\cdots$ run from $2$ to $n~(\equiv D-1)$,
\begin{equation} \label{mainlinear} 2\delta R_{i j}=\left(\Delta_L
h\right)_{i j}= -\,2nA\,h_{i j}\,,\end{equation} with
\begin{eqnarray} \left(\Delta_L h\right)_{i j} &=&
\frac{1}{r^2}\,\tilde{\Delta}_L h_{i j} -f \left(\frac{d^2}{d r^2}
+\frac{4}{r^2}\right)h_{i j} \nn \\
&{}&-\frac{r f^\p+(n-5)f}{r}\,\frac{d}{d r}\,h_{i j}+\frac{1}{f}\,
\frac{d^2}{d t^2}\,h_{i j} \,,\nn \\
A&=& \frac{C}{1-2C\alpha}\,\left[1-\frac{C \alpha
(3n^2-3n+2)}{n(n-3)}\right], \end{eqnarray} where
$\tilde{\Delta}_L h_{i j}$ is the Lichnerowicz operator acting on
${\cal M}$. We look for unstable tensor modes of the form
\begin{equation} h_{i j}(x) = r^2\,\phi(r)\, e^{\omega t}\,\wth_{i
j} (\tilde{x}),
\end{equation} where $(\tilde{x})$ are coordinates on ${\cal
M}$. It is convenient to assume that \begin{equation}
\left(\widetilde{\Delta}_L \wth\right)_{i j}=\lambda\, \wth_{i
j},\end{equation} where $\lambda$ is the eigenvalue of the
Lichnerowicz operator. Equation~(\ref{mainlinear}) then takes the
form of a Sturm-Liouville type problem~\cite{Gibbons02a}
\begin{eqnarray} &&-\,f\frac{d}{dr}\left(f\,
r^{n-1}\,\frac{d\phi(r)}{dr}\right)\nonumber \\
&&~~ -\,\frac{2f}{r^2}\,\left((n-2)f -\frac{\lambda}{2}
+r\,f^\p-nA\,r^2\right)\phi(r)\nonumber \\
&&~~=\omega^2\,\phi(r) \,r^{n-1}\,.\end{eqnarray} In terms of
Regge-Wheeler type coordinates $dr\equiv f\,dr_*$ and
$\phi(r)\equiv \Phi\,r^{-(n-1)/2}$, this takes the form
\begin{equation} \frac{d^2\Phi}{dr_*^2}- V\left(r(r_*)\right)\Phi
=\omega^2 \Phi,
\end{equation} where \begin{eqnarray}
\label{mainpoten}
V(r)&=&\frac{\lambda\,f}{r^2}+\frac{\left[(n-2)(n-10)-1\right]
f^2}{4r^2} + \frac{(n-5)ff^\p}{2r}\nn \\
&{}& ~~~~~~~~~~~~~~~~~~~~~~~~~~~~~~~~~~ + 2nA\,f .\end{eqnarray} A
requirement of finite energy is equivalent to the normalization
condition that $\int \Phi^2\,dr_*=\int f(r)^{-1}\,\Phi^2\,dr$ is
finite. Usually, the stability of a potential depends on the
eigenvalues of the Lichnerowicz operator, ensuring that the
potential is positive and bounded from below.

\subsection{Stability of massless state in general relativity}

The Schwarzschild black hole is stable in four
dimensions~\cite{Vishves70a}. It is also learnt that a spherically
symmetric AdS$_4$ black hole is stable against small
electromagnetic and gravitational perturbations~\cite{Cardoso01a}.
The issue of instability of tensor modes, however, arises only for
$n+1>4$ since there are no tensor harmonics on $S^2$ or $H^2$.

For a vanishing cosmological constant ($\Lambda=0$), so that
$A=0$, the spacetime metric is flat if the base is a sphere, and
if the base is not a sphere then the full spacetime is a cone
which is singular at the origin. Thus, for $\Lambda=0$, we must
take $k=1$, so that $f(r)\to 1$ asymptotically and $r=r_*$. Then
the asymptotic potential is
\begin{equation} \label{inftypoten}
V_\infty(r)=\frac{4\lambda+(n-2)(n-10)-1}{4\,r^2}
=\frac{(2\nu-1)(2\nu+1)}{4\,r^2}\,. \end{equation}  This potential
is non-negative only if
\begin{eqnarray} \label{lambdarel.} \lambda &\geqslant &
\frac{1+(n-2)(10-n)}{4}\,,\nn
\\ or \quad \nu &=&\frac{1}{2}\,\sqrt{(n-2)(n-10)+4\lambda}\geqslant \frac{1}{2}, \end{eqnarray} where $V_\infty (r)=0$ if equality
holds. Stability of a potential requires that there are no bound
negative energy states or no growing modes with $\omega>0$ at the
horizon, if the latter exists. A requirement that $\lambda>0$
indeed constraints the spacetime dimensions $D=(n+1)$ to
$n\leqslant 10$. The asymptotic solution for $\Phi(r)$ that decays
as $r$ goes to $+\infty$ is \begin{eqnarray} \Phi_\infty &=&
C_1\,\sqrt{r}\, J_\nu\left(\sqrt{-\omega^2}\,r\right)+
C_2\,\sqrt{r}\,~  Y_\nu \left(\sqrt{-\omega^2}\,r\right)\nn \\
&=& \mbox{Re}\left[\sqrt{r}\,K_\nu(\omega r)\right].
\end{eqnarray} For small $r$ and a real positive $\nu$,
$\Phi_\infty(r)$ behaves as $\sim r^{-\nu+1/2}$. This solution is
divergent but normalizable if $1>\nu >1/2$, and divergent and
non-normalizable if $\nu\geqslant 1$. For $n\leqslant 10$, an
imaginary $\nu$ does not exist with a positive potential, while,
for $n>10$, the eigenvalue $\lambda$ may take a negative value, so
$\nu>1$, but this solution is always non-normalizable.

For $\Lambda<0$, the potential~(\ref{inftypoten}) is modified to
\begin{eqnarray} \label{fullpotenk=1}
V(r)&=&\left(\frac{4\lambda+(n-2)(n-10)-1}{4\,r^2}
+\frac{(n-1)(n+1)}{4\,l^2}\right)\nn \\
&{}& ~~~~~~~~~~~~~~~~~~~~~~\times
\left(1+\frac{r^2}{l^2}\right)\nn \\
V(x) &=&\frac{1}{l^2} \Bigg(\frac{4(\lambda-n+1)+(n-3)(n-5)}{4\,
x} \nn \\
&{}& ~~~~~~~~~~~~~~ +\,
\frac{(n-1)(n+1)}{4}\Bigg)\left(1+x\right), \end{eqnarray} where
$x\equiv r^2/l^2$. This is always positive and bounded from below
in satisfying Eq.~(\ref{lambdarel.}). The Schr\"odinger equation
may be expressed as a hypergeometric equation
\begin{eqnarray} \label{Schro} &{}& x(x+1)^2\,\Phi^{\p\p}(x)+
\frac{(x+1)(3x+1)}{2}\,\Phi^\p(x)\nn \\
&{}&~~ -\, \frac{x+1}{4}\times
\Bigg(\frac{4\lambda+(n-2)(n-10)-1}{4x}\nn \\
&{}& ~~~~~+\,
\frac{(n-1)(n+1)}{4}+\frac{\omega^2\,l^2}{(x+1)}\Bigg)\Phi(x)=0,
\end{eqnarray} with no singularities in the range $0<x<\infty$. To
simplify~(\ref{Schro}) one can make the following substitutions:
\begin{eqnarray} 2\nu &=&\sqrt{4\lambda+(n-2)(n-10)},
\quad \sigma=-\,\frac{i\omega\,l}{2},\nn \\
a&=&\sigma+\frac{2\nu-(n-2)}{4}\,, \quad
b=\sigma+\frac{2\nu+(n+2)}{4}\,, \nn \\
&{}&~~~~~~~~~~~~~~~~~ c=\nu+1\,. \end{eqnarray} Under the
decomposition $\Phi(x)=x^{(1+2\nu)/4}\, (1+x)^\sigma\,\varphi(x)$,
two independent solutions of Eq.~(\ref{Schro}) are found to be
\begin{equation}
    \varphi(x) = \left\{\begin{array}{l}
    {}_2F_1(a,\, b;\, c;\, -\,x), \\
    x^{1-c}\,~{}_2F_1(a-c+1,\, b-c+1;\, 2-c;\, -\,x),
    \end{array} \right.\
    \end{equation}
where ${}_2F_1(a, b; c;x)$ is the hypergeometric function, and the
parameter $c\neq 0, -1, -2, -3, \cdots$ for the first solution,
and $c\neq 2, 3, 4,\cdots$ for the second solution. A general
solution of Eq.~(\ref{Schro}) is given by a linear combination of
\begin{eqnarray} \Phi_\pm (x)&=& x^{(1\pm
2\nu)/4}\,\left(1+x\right)^{-\,i\omega l/2}
{}_2F_1\Bigg(\frac{\pm\, 2\nu-(n-2)}{4}\nonumber \\
&{}&-\frac{i\omega l}{2},\, \frac{\pm\,
2\nu+(n+2)}{4}-\frac{i\omega l}{2};\,
\pm\,\nu+1;\,-\,x\Bigg).\nonumber \\
\end{eqnarray} A stability check may be
needed in both limits $x\to 0$ and $x\to \infty$. This is the case
studied by Gibbons and Hartnoll~\cite{Gibbons02a} in some detail.
For consideration of the $\alpha>0$ case in the next section, we
briefly discuss first some important features of the solution with
$\alpha^\prime=0$.

To study the asymptotic behavior, one uses the following relation
of the hypergeometric functions~\cite{Askey} \begin{eqnarray}
{}_2F_1(a,b;c;x)&=&h_1(-x)^{-a}
{}_2F_1\left(a, a-c+1; a-b+1;\frac{1}{x}\right)\nn \\
&+&h_2\,(-x)^{-b}\,{}_2F_1\left(b,\,b-c+1; b-a+1;
\frac{1}{x}\right), \nonumber \\
\label{infinite} \end{eqnarray}
where
\begin{equation}
h_1=\frac{\Gamma(c)\Gamma(b-a)}{\Gamma(c-a)\Gamma(b)}\,, \quad
h_2=\frac{\Gamma(c)\Gamma(a-b)}{\Gamma(c-b)\Gamma(a)}\,.
\end{equation}
As $x\to \infty$, the asymptotic solution is given by
\begin{eqnarray} && {}_2F_1(a, b; c; x)=h_1 (-x)^{-a}
\left(1+\frac{a(a-c+1)}{a-b+1} \frac{1}{x}+\cdots\right)\nn \\
&&~~~~~~~~~+\,h_2 (-x)^{-b} \left(1+\frac{b(b-c+1)}{b-a+1}
\frac{1}{x}+\cdots\right).\end{eqnarray} The boundedness requires
that as $r\to\infty$ the function $\Phi(r)$ must go as ${\cal
O}\left(r^{(n-1)/2}\right)$ or a lower power of $r$. A solution
that has this behavior is given by a linear combination of
$\Phi_{\pm}$ such that the term $(x)^{-b}$ of the two solutions
cancels. Hence,
\begin{eqnarray} \label{thirdsol} \Phi_3 (x)&=&
x^{-(n+1)/4+i\omega\,l/2}\,\left(1+x\right)^{-\,i\omega\,l/2} \nn
\\&\times & {}_2F_1\Bigg(\frac{2\nu+n+2}{4}-\frac{i\omega\,l}{2},
\, \frac{-\, 2\nu+n+2}{4}-\frac{i\omega\,l}{2};\nn \\
&{}&~~~~~~~~~~~~~~~~~ \,\frac{n+2}{2};\,-\,\frac{1}{x}
\Bigg)\,.\end{eqnarray} To evaluate the behavior as $r\to 0$, one
can apply Eq.~(\ref{infinite}) to Eq.~(\ref{thirdsol}), replacing
$1/x$ by $x$ for ${}_2F_1$. The leading behavior of the solution
as $x\to 0$ is given by ${\cal
O}\left(x^{-(n+1+2\nu-n-2)/4}\right) \sim {\cal
O}\left(r^{(1-2\nu)/2}\right)$. The range of convergence for
$c>a+b$ is $|x|<1$ and $x=1$, which implies that
$0>-\,i\omega\,l$. That is, a situation that the hypergeometric
series terminates for some special (positive) values of $\omega$
simply does not arise in the above case. As a result, the massless
configurations where ${\cal M}$ has constant positive ($k=+1$)
curvature are stable under tensor perturbations.

\subsection{Instability of massless topological black holes}

\begin{figure}[ht]
\begin{center}
\epsfig{figure=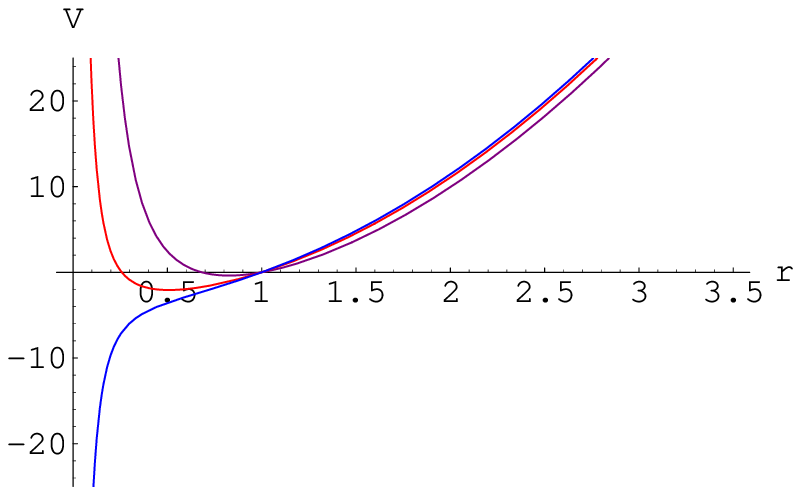,height=3.5cm,width=6.5cm}
\end{center}
\begin{center}
\epsfig{figure=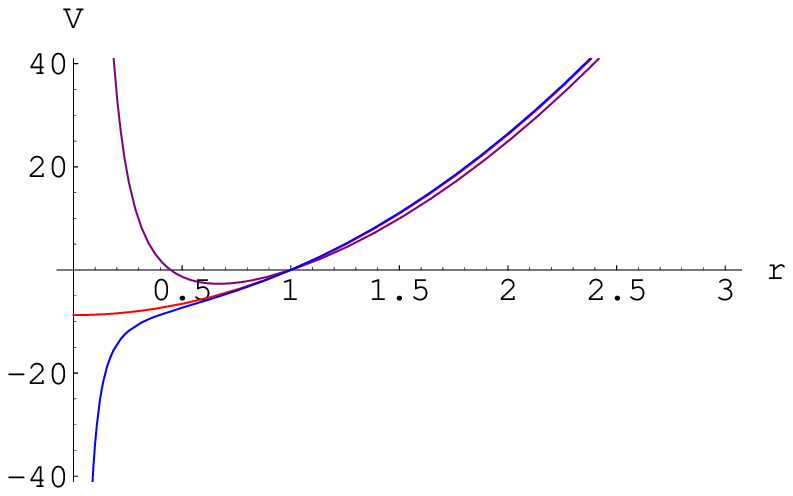,height=3.5cm,width=6.5cm}
\end{center}
\begin{center}
\epsfig{figure=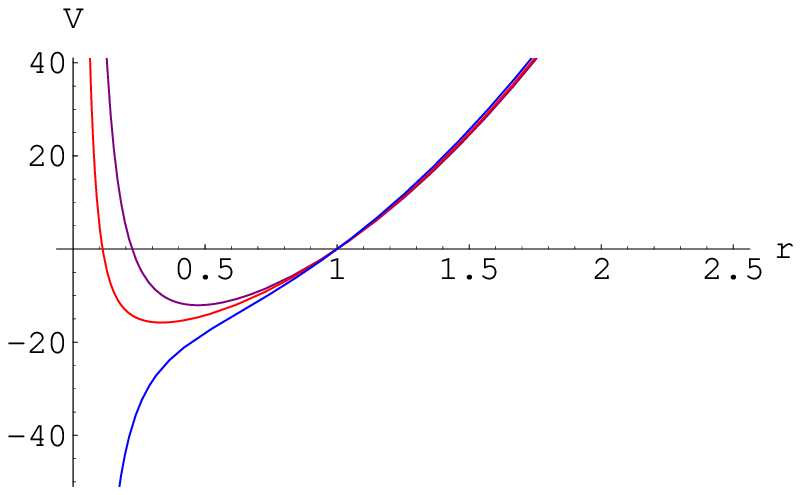,height=3.5cm,width=6.5cm}
\end{center}
\caption{The potential as a function of the horizon $r_+$ at fixed
$l=1$: (top panel) $n=4$ and $\lambda=-\,5, -\,14/4, -\,3$ (top to
bottom); (middle panel) $n=6$ and $\lambda=-\,6, -\,17/4, -\,4$
(top to bottom); (bottom panel) $n=9$ and
$\lambda=-\,3,~-\,9/4,~-\,1$ (top to bottom). The above potential
will be bounded and positive only for $\lambda=5-3n$, and it is
always unbounded for $\lambda$ satisfying Eq.~(\ref{mainconstr}).}
\label{figure9}
\end{figure}

A dynamical instability of topological black holes in Einstein
gravity, under tensor perturbations, was studied in
Ref.~\cite{Gibbons02a}. So we will be brief in our analysis of
this particular case, but we shall review some results reported
earlier as a quantitative difference arises. For $\alpha=0$, and
$k=-1$, $\mu=0$, the metric solution is
\begin{equation} \label{bhmetric1} f(r)=-1+\frac{r^2}{l^2}
\,.\end{equation} The gravitational potential then takes the form
\begin{eqnarray}\label{potential4}
V(x)&=&\frac{1}{l^2}\Bigg(\frac{4(n-1+\lambda)-(n-3)(n-5)}{4\,x}\nn
\\ &{}&~~~~~~ +\,\frac{(n-1)(n+1)}{4}\Bigg)\left(x-1\right),
\end{eqnarray} where
$x\equiv r^2/l^2$. One may easily check that this potential is
positive for all values of $x$ only if \begin{equation}
\lambda=5-3n\equiv \tilde{\lambda}\,.\end{equation} This is a
special situation. Since the Lichnerowicz spectrum is bounded
below by $\lambda_{min}=-\,2(n-1)$, the spacetime dimension
$n+1=4$ is on the borderline for which
$\tilde{\lambda}=\lambda_{min}$. For $n+1\geqslant 5$,
$\tilde{\lambda}$ is smaller than $\lambda_{min}$, and the
potential can be negative and unbounded. More precisely, as the
plots in Fig.~\ref{figure9} show, the potential is negative but
bounded from below when \begin{equation} \label{critical}
\lambda<\lambda_{crit}=\frac{(n-2)(n-10)-1}{4}\,,\end{equation}
and unbounded when $\lambda \geqslant \lambda_{crit}$; the latter
might signal instability of a massless topological black hole.

To gather more information, we need to solve the full differential
equation. The Schr\"odinger equation, under the
potential~(\ref{potential4}), may be expressed as
\begin{eqnarray} \label{HyperGk=-1} &&
x(x-1)\,\Phi^{\p\p}(x)+\frac{x+1}{2}\,\Phi^\p(x)-\frac{1}{4}\,
\Bigg(\frac{(n-1)(n+1)}{4}\nn \\
&& +\frac{4\lambda+1-(n-2)(n-10)}{4x}+\frac{\omega^2\,l^2}{4}\,
\frac{1}{x-1}\Bigg)\Phi(x)=0\,. \nn \\
&&~~~~~~~~~~~~~~~~~~ \end{eqnarray} Let us make the following
substitutions:
\begin{eqnarray} \label{substitute} 2\tnu
&=&\sqrt{4(2-\lambda)+(n-2)(n-10)}\,,
\quad \tsigma=-\,\frac{\omega\,l}{2},\nn \\
a&=&\tsigma+\frac{2\tnu+(n+2)}{4}\,, \quad
b=\tsigma+\frac{2\tnu-(n-2)}{4}\,, \nn \\
&{}&~~~~~~~~~~~~~~~~~ c=\tnu+1,
\end{eqnarray} and decompose the
harmonic function as $\Phi(x)=x^{(3+2\tnu)/4}\,
{(x-1)}^{\tsigma}\,\varphi(x)$. Then two independent solutions of
Eq.~(\ref{HyperGk=-1}) are easily written as
\begin{equation}
    \varphi(x) = \left\{\begin{array}{l}
    {}_2F_1(a, b; c; x),\\
    x^{1-c} {}_2F_1(a-c+1, b-c+1; 2-c; x).
    \end{array} \right.\
    \end{equation}
The general solution of Eq.~(\ref{HyperGk=-1}) is given by a
linear combination of the following two solutions:
\begin{eqnarray} \label{k-1sol} &{}& \Phi_\pm (x)=x^{(3\pm
2\tnu)/4}\,\left(x-1\right)^{-\,\omega\,l/2}\times \nn \\
&{}& {}_2F_1\Bigg(\frac{\pm\,
2\tnu+(n+2)}{4}-\frac{\omega\,l}{2},\, \frac{\pm\,
2\tnu-(n-2)}{4}-\frac{\omega\,l}{2};\nn \\
&{}&~~~~~~~~~~~~~~~~~~~~~~~~\,
\pm\,\tnu+1;\,\,x\Bigg).\end{eqnarray} The boundedness requires
that as $r\to\infty$ the function $\Phi(r)$ must go as ${\cal
O}\left(r^{(n-1)/2}\right)$ or a lower power of $r$. The solution
that has this behavior is given by a linear combination of
$\Phi_{\pm}$ such that the term $(-x)^{-b}$ of two solutions
cancel, and hence the asymptotic behavior goes like ${\cal
O}\left(x^{(3\pm 2\tnu\mp 2\tnu -n-2)/4}\right)\sim {\cal
O}\left(r^{-(n-1)/2}\right)$. The solution that is well behaved as
$r\to \infty$ reads
\begin{eqnarray} \Phi_3 (x)&=&x^{(2\omega l
-n+1)/4}\left(x-1\right)^{-\,\omega {l}/2}
{}_2F_1\Bigg(\frac{2\tnu+n+2}{4}-\frac{\omega {l}}{2},\nonumber \\
&{}& \frac{-2\tnu+n+2}{4}-\frac{\omega\,l}{2}; \frac{n+2}{2};
\frac{1}{x}\Bigg).
\end{eqnarray}
We can find a new solution of the hypergeometric equation by using
the relation \begin{eqnarray}{}_2F_1\left(a,\, b;\,c;\,x\right)&=&
k_1 ~{}_2F_1\left(a,\,b;\,a+b+1-c;\,1-x\right)\nn \\
&{}& +\, k_2\left(1-x\right)^{c-a-b} {}_2F_1 \Big(c-b, c-a;
\nonumber\\
&{}& 1+c-a-b; 1-x\Big),\end{eqnarray} where
\begin{equation}
k_1=\frac{\Gamma(c)\Gamma(c-a-b)}{\Gamma(c-a)\Gamma(c-b)}\,,\quad
k_2=\frac{\Gamma(c)\Gamma(-c+a+b)}{\Gamma(a)\Gamma(b)}\,.
\end{equation} For a generic solution with $k_1, k_2\neq 0$, the
dominant term as $x\to 1$ is ${\cal
O}\left((x-1)^{-\omega{l}/2}\right)$. This solution is not square
integrable at $x=1$. In the special case, with $k_1=0$, which is
accomplished by choosing $c=a$ or $c=b$, one finds a solution with
better behavior at $x=1$. In this case, the leading term as $x\to
1$ is
\begin{equation}
(x-1)^{-\omega\,l/2}\,\left(\frac{x-1}{x}\right)^{\omega\,l}=
{\cal O}\left((x-1)^{\omega\,l/2}\right).
\end{equation} This
solution is normalizable and bounded at the horizon for
$\omega>0$. However, the condition that either $(c-a)$ or $(c-b)$
is zero would be that
\begin{equation} \omega =\left(\frac{2\tnu-(n+2)}{2\,l}\right)>0
\,.\end{equation} This holds if $2\tnu>n+2$ or equivalently,
$\lambda_{min} <(6-4n)$ [cf., Eq.~(\ref{substitute})], which is
not allowed since the Lichnerowicz spectrum is bounded below by
$\lambda_{min}= -\,2(n-1)$. In the following range of the
eigenvalues, which are allowed in the theory such that $2\tnu
> 1$:
\begin{equation} \label{mainconstr}
\frac{(n-2)(n-10)-1}{4}\leqslant \lambda <
\frac{(n-2)(n-10)+8}{4},
\end{equation} we still have $\lambda>\lambda_{min}$,
the gravitational potential is always unbounded from below.
In~\cite{Gibbons02a} a massless $k=-1$ background was expected to
be classically gravitationally stable if the
$\lambda>\lambda_{crit}$. Whether this interpretation is correct
is not clear {\it a priori} since one would not expect to have a
classically stable background if the potential is negative and
unbounded from below.

In~\cite{Aros02a}, by taking the massless $k=-1$ black hole itself
as a ground state, an analytic result for the quasinormal modes of
a scalar perturbation was presented for $n\geqslant 3$. Here we
emphasize that a metric background with $k=-1$ and $\mu=0$ may be
unstable under tensor perturbations. Let us give some specific
examples. Instability of the $\mu=0$, $k=-1$ spacetime may arise
when
\begin{eqnarray}  &{}& n=4: \quad -\,\frac{13}{4}\leqslant
\lambda
<-\,1, \nn \\
 &{}& n=6: \quad -\,\frac{17}{4}\leqslant \lambda <-\,2,\nn \\
 &{}& n=9: \quad -\,2\leqslant \lambda <\frac{1}{4}.
 \nonumber
\end{eqnarray} In these ranges, the gravitational potential is
negative and unbounded from below. Although a negative potential
does not necessarily imply instability of a background, an
unbounded potential certainly signals instability of the
background metric with $\mu=0$ and $k=-1$.

\subsection{Stability of negative mass extremal state}

When $\alpha^\prime=0$, the extremal solution is given by
\begin{equation} \label{extremalMetric} f(r)=-1+\frac{r^2}{l^2}+
B\left(\frac{l}{r}\right)^{n-2}\,, \quad
B=2\sqrt{\frac{(n-2)^{n-2}}{n^n}}\,. \end{equation} The
gravitational potential therefore reads \begin{eqnarray}
\label{extremal1+n}
V(r)&=&\Bigg[\frac{4(n-1+\lambda)-(n-3)(n-5)}{4{r}^2}
+\frac{(n-1)(n+1)}{4{l}^2}\nonumber \\
&{}&-\,\frac{B(n-1)^2 l^{n-2}}{4
r^n}\Bigg]\left[-1+\frac{r^2}{l^2}
+B\left(\frac{l}{r}\right)^{n-2}\right].
\end{eqnarray} The spacetime one is allowed to take is
$r>r_{extr}=l\,\sqrt{(n-2)/n}$.

\begin{figure}[ht]
\begin{center}
\epsfig{figure=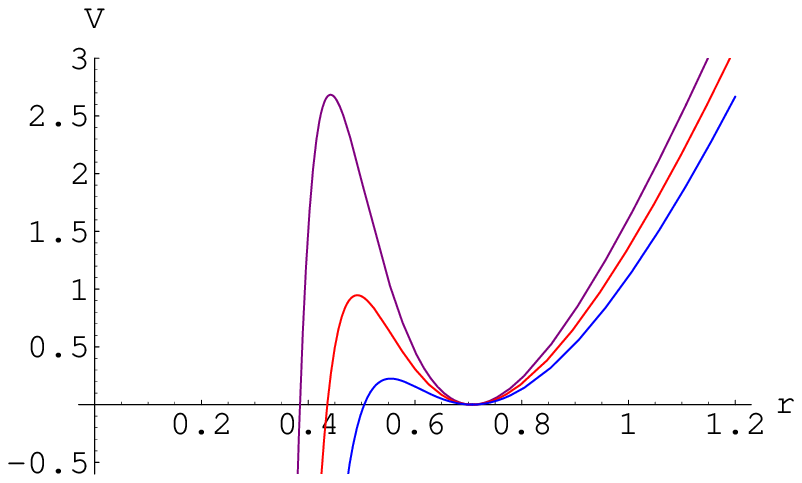,height=3.5cm,width=6.5cm}
\end{center}
\begin{center}
\epsfig{figure=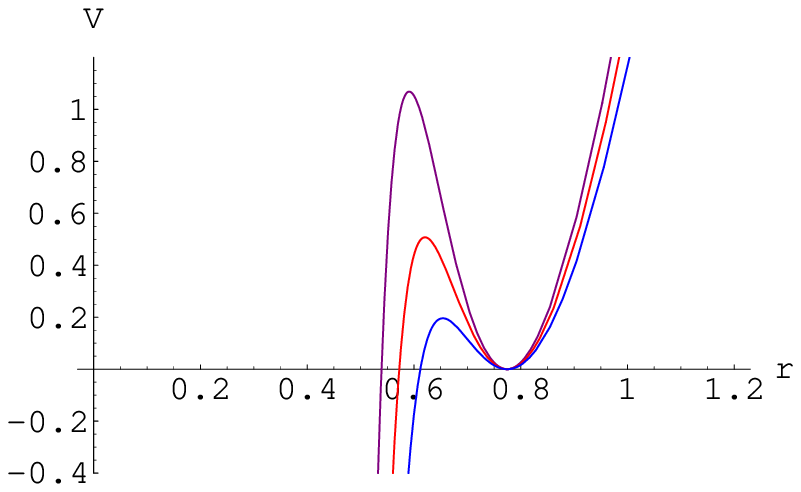,height=3.5cm,width=6.5cm}
\end{center}
\begin{center}
\epsfig{figure=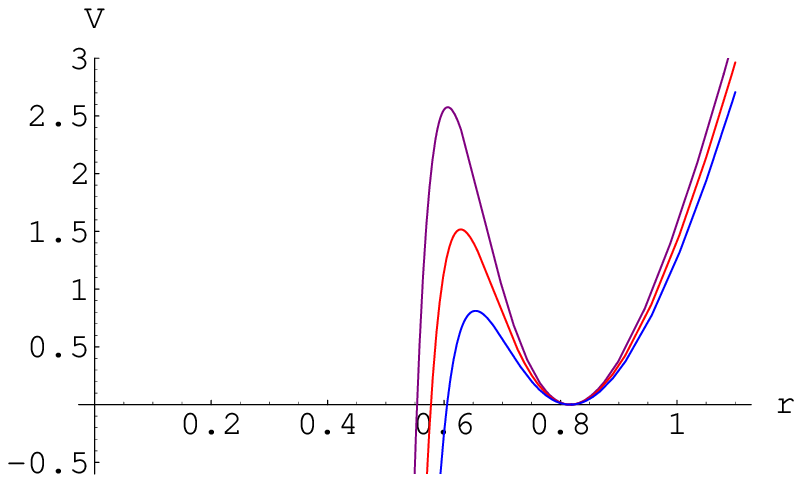,height=3.5cm,width=6.5cm}
\end{center}
\caption{The potential defined for a negative mass extremal
background with $\alpha=0$. The values are chosen, from top to
bottom, at (top panel) $n=4$ and $\lambda=0, -1$, and $-2$;
(middle panel) $n=5$ and $\lambda=-1, -2$, and $-3$; (bottom
panel) $n=6$ and $\lambda=-2, -3$, and $-4$.} \label{figure10}
\end{figure}


For $n=4$, the metric function~(\ref{extremalMetric}) takes the
form \begin{equation} \label{4dextremalM}
f(x)=\frac{(2x-1)^2}{4x}\,, \end{equation} where $x\equiv r^2/l^2$
is a dimensionless scale. As there is no cosmological horizon for
an extremal solution, the perturbation extends to $x\to \infty$.
The potential
 \begin{equation}
V(x)=\frac{1}{l^2}\,\left(\frac{4(3+\lambda)+1}{4\,x}+\frac{15}{4}
-\frac{9}{16\,x^2}\right) \left(\frac{(2x-1)^2}{4\,x}\right)\,
\end{equation} vanishes at the extremal horizon $2x=1$, but it
is everywhere positive outside the extremal horizon. Under this
potential, the Schr\"odinger equation takes the form
\begin{eqnarray} \label{extremal2+n}
&&\frac{(2x-1)^2}{8x^2}
\Bigg[2x(2x-1)^2\,\Phi^{\p\p}(x)+(4x^2-1)\,\Phi^\p(x)\nonumber \\
&&~-\,2x
\left(\frac{4(3+\lambda)+1}{4{x}}+\frac{15}{4}-\frac{9}{16
x^2}\right)\Phi(x)\Bigg]\nonumber\\
&&~~~~~~ -\omega^2 l^2\,\Phi(x)=0.
\end{eqnarray} The condition of
boundedness is automatic, because it simply requires that
$\Phi(x)$ is bounded at $2x=1$, where $f(x)=0$. The finite energy
condition will be that $\Phi(x)$ goes to zero on the extremal
horizon, because the zero of $f(x)$ is simple. We can get some
insights into the stability of the solutions also by inspecting
the plots for the gravitational potential.

The hypergeometric equation~(\ref{extremal2+n}) can be solved
exactly only for the $\omega=0$ modes: \begin{eqnarray}
\Phi(x)&=&c_1\,(2x-1)^{-\,\gamma/4}\,x^{3/4}{}_2F_1(a,b;1;2x)\nn
\\
&{}& +\,c_2\, (2x-1)^{-\,\gamma/4}\,
x^{3/4}{}_2F_1(a,b;1,2x)\nn \\
&{}&\, \times \int
\frac{(2x-1)^{-1+\gamma/2}}{x\,\left[{}_2F_1(a,b;1;2x)\right]^2}\,dx
,\end{eqnarray} where \begin{equation}
\gamma=\sqrt{2\lambda+32}\,,\quad a=\frac{6-\gamma}{4}\,,\quad
b=\frac{-\,(2+\gamma)}{4}\,.\end{equation} This solution converges
at $2x=1$ for any eigenvalue $\lambda>\lambda_{crit}$, and also in
the range~(\ref{mainconstr}), and it is also normalizable there.
The situation is similar in the $n>4$ case. Now for the
eigenvalues $\lambda$ in the range given by
Eq.~(\ref{mainconstr}), where one might expect instability to
arise for a massless background, the potential is bounded and
positive. Therefore, in the spacetime region $r\geqslant
r_{extr}$, the potential is always positive, tending to zero at
the extremal state. This means that the extremal ground state may
be stable under tensor perturbation.


For $n=6$, the metric function~(\ref{extremalMetric}) takes the
form \begin{equation} f(x)=\frac{(3x+1)(3x-2)^2}{27\,x^2}\,,
\end{equation} and the corresponding potential is \begin{eqnarray}
V(x)&=&\frac{1}{l^2}\,\left(\frac{4\lambda+17}{4x}+\frac{35}{4}
-\frac{25}{27\,x^2}\right)\nn \\
&{}&~~~~~~\times \left( \frac{(3x+1)(3x-2)^2}{27\,x^2}\right)\,.
\end{eqnarray}
This does not cover the small $x$ region; for example, when
$\lambda=-13/4$, the potential is not bounded from below when
$x<0.27$. That is, the spacetime region $r<r_{e}$ may have an
internal infinity, where the potential is not well behaved, and
the reference spacetime is incomplete in Einstein gravity,
especially for the $k=-1$ case. This is clearly seen from the
plots of gravitational potential $V(r)$ vs horizon position in
Fig.~\ref{figure10}. This problem may be resolved once the
background metric and hence the potential receives higher
derivative curvature corrections.

\section{Gauss-Bonnet black holes and Gravitational Stability}

In this section, we study the stability of metric backgrounds with
a non trivial GB coupling. With $\mu=0$, the AdS vacuum solution
is given by \begin{equation} \label{mu=0}
f(r)=k+\frac{r^2}{2\alpha}\left(1\mp \sqrt{1-
\frac{4\alpha}{l^2}}\,\right)\equiv k+C\,r^2\,. \end{equation} In
this case, for any $k$, the spacetime metric has constant
curvature. However, for $k=-1$, we take the negative mass extremal
state as a ground state, which reads in five spacetime dimensions
as
\begin{equation} \label{GBExtremal}
f(r)=-1+\frac{r^2}{2\alpha}\left[1 - \sqrt{\left(1 -
\frac{4\alpha}{l^2}\right)\left(1-\frac{\alpha
l^2}{r^4}\right)}\right]\,.\end{equation} The metric spacetime is
only asymptotically locally AdS, which can be easily seen by
allowing a particular coupling. The requirement that $4\alpha <
l^2$ is complementary to the condition $r^2 \geqslant l^2/2$.

\subsection{Stability of a massless Gauss-Bonnet black hole}

\begin{figure}[ht]
\begin{center}
\epsfig{figure=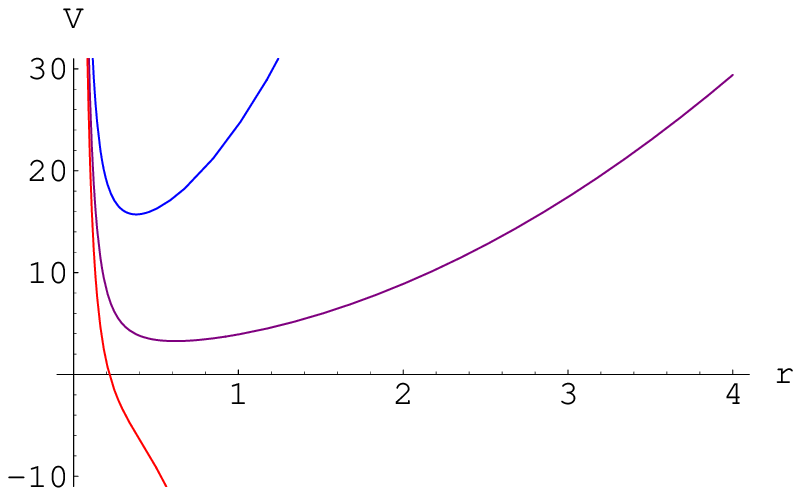,height=3.5cm,width=6.5cm}
\end{center}
\begin{center}
\epsfig{figure=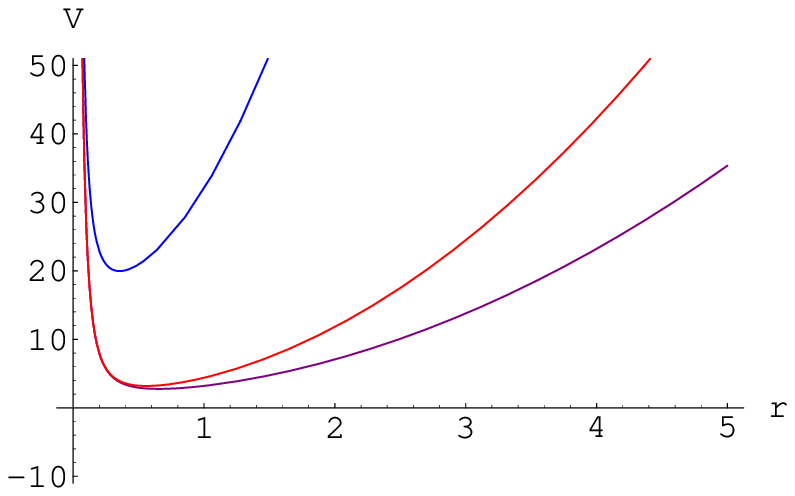,height=3.5cm,width=6.5cm}
\end{center}
\begin{center}
\epsfig{figure=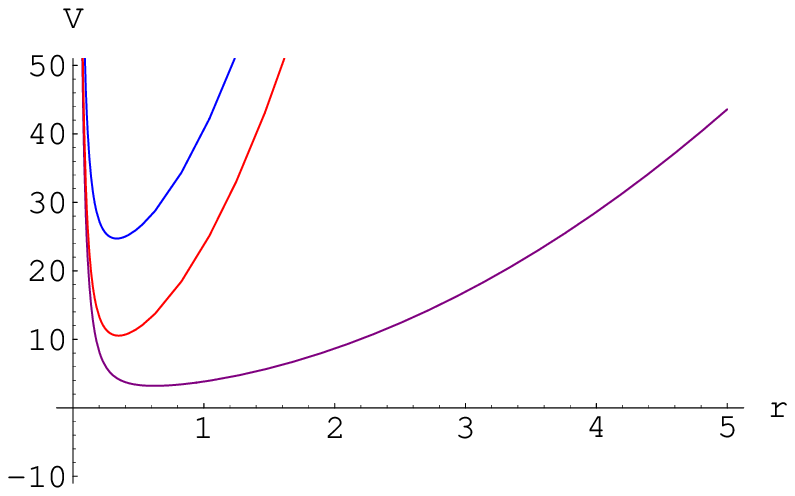,height=3.5cm,width=6.5cm}
\end{center}
\caption{The potential vs horizon radius for $k=+1$. The values
are fixed at $n=7$, $\lambda=17/4$, $\alpha=2/11$ (upper plot);
$n=8$, $\lambda=14/4$, $\alpha=2/9$ (middle plot); $n=9$,
$\lambda=9/4$, $\alpha=1/4$ (lower plot). The uppermost curve
corresponds to the $\alpha=0$ case, and the two other curves
correspond to $\mp$ signs in Eq.~(\ref{k=1bg}): $-$ and $+$ (upper
and lower) in the uppermost plot, and the reverse in the lower two
plots.} \label{figure11}
\end{figure}

As shown in the previous section, for $\alpha=0$, the AdS black
holes with $\mu=0$ and $k=+1$ are stable under tensor
perturbations. Here we extend this analysis for a non trivial GB
coupling. The background metric function, with $\alpha>0$ and
$k=+1$, is
\begin{equation} \label{k=1bg} f(r)=1+C\,r^2 \,, \quad
\frac{1}{C}\equiv\frac{2\alpha}{1\mp
\sqrt{1+\frac{8\alpha\Lambda}{n(n-1)}}}
>0\,.\end{equation}
Let us introduce a dimensionless scale $x=Cr^2$; then the
corresponding Schr\"odinger equation may be expressed as a
hypergeometric equation \begin{eqnarray} \label{GBSchwarz} &{}&
x(x+1)^2\,\Phi^{\p\p}(x)+
\frac{(x+1)(3x+1)}{2}\,\Phi^\p(x)\nn \\
&{}& -\, \frac{x+1}{4}
\Bigg(\frac{4\lambda+(n-2)(n-10)-1}{4x}\nn \\
&{}&+\, \frac{n^2-8n-1+d}{4}+
\frac{\omega^2}{(x+1)C}\Bigg)\Phi(x)=0, \end{eqnarray} where
\begin{eqnarray}
\label{constant=d} && d= \frac{8nC}{(1-2C\alpha)}
\frac{2\alpha}{1\mp \left(1-2C\alpha\right)}
\left[1-\frac{C\alpha(3n^2-3n+2)}{n(n-3)}\right].\nonumber \\
\end{eqnarray} We
shall choose the minus sign of the $\mp$ in Eq.~(\ref{k=1bg}) or
in Eq.~(\ref{constant=d}), because only this branch gives a stable
solution in all dimensions (Fig.~\ref{figure11}). Another reason
for choosing the negative sign is that, for this branch, in the
limit $\alpha\to 0$, the black hole solutions reduce to that of
Einstein gravity.


The hypergeometric equation~(\ref{GBSchwarz}) may be solved
exactly, and two independent solutions are \begin{eqnarray}
\Phi_\pm (x)&=&x^{(1\pm 2\nu)/4}\left(1+x\right)^{-\,i\omega
/{2\sqrt{C}}}{}_2F_1\Bigg(\frac{\pm\, 2\nu \mp 2\beta
+2}{4}\nonumber \\
&{}&-\,\frac{i\omega}{2\sqrt{C}}, \frac{\pm\, 2\nu \pm 2\beta
+2}{4}-
\frac{i\omega}{2\sqrt{C}}; \pm\,\nu+1; -\,x\Bigg), \nonumber \\
\end{eqnarray}
where $\nu=\frac{1}{2}\,\sqrt{4\lambda+(n-2)(n-10)}$ and
$\beta=\frac{1}{2}\,\sqrt{n^2-8n+d}$. To determine the conditions
for stability, with real $\nu$ and $\beta$, we need to satisfy
\begin{equation} \lambda \geqslant \lambda_{crit} =
\frac{1+(n-2)(10-n)}{4}\,,\quad d\geqslant 8n-n^2+1
\,.\end{equation} The first constraint is the same as in the
$\alpha=0$ case, where $d=8n$, so $2\beta=n$. We may derive a more
useful constraint for $C\alpha$ from the condition $d\geqslant
(8n-n^2+1)$, so $2\beta\geqslant 1$, which implies that
\begin{equation} \label{bound} C\alpha \leqslant \frac{(n-3)(n-1)}{2(n^2+11)}\,.\end{equation} Using
Eq.~(\ref{k=1bg}), along with $\Lambda=-\,n(n-1)/(2l^2)$, we find
\begin{equation} \label{newconstr} \frac{\alpha}{l^2}\leqslant \frac{(n-3)(n-1)(n^2+4n+19)}{4(n^2+11)^2}\,. \end{equation} For
$n+1=5$, one finds $\alpha/l^2\leqslant 17/324$. The bound
$\alpha/l^2< 17/324$ required here to keep the potential
non-negative at the linearized level already puts a stronger bound
to $\alpha/l^2$ than needed for thermodynamic (or dynamical)
stability, namely, $\alpha/l^2\leqslant 1/12$.

By the same argument as for the pure AdS case with $\alpha=0$, the
solution that is well behaved as $x\to \infty$ is given by the
linear combination of $\Phi_\pm$, which reads \begin{eqnarray}
\Phi_3 (x)&=&
x^{-(2\beta+1)/4+i\omega/{2\sqrt{C}}}\,\left(1+x\right)^{-\,i\omega
/{2\sqrt{C}}} \nn \\
&{}& {}_2F_1\Bigg(\frac{2\beta + 2\nu +2}{4}-\frac{i\omega
}{2\sqrt{C}},\, \frac{2\beta - 2\nu +2}{4}\nn \\ &{}& ~~ -\,
\frac{i\omega}{2\sqrt{C}};\,\beta+1;\,-\,\frac{1}{x}\Bigg)\,.
\end{eqnarray}
In the $\alpha=0$ case, the leading behavior as $r\to \infty$ goes
as ${\cal O}\left(r^{-(n+1)/2}\right)$, while as $r\to 0$ it goes
as ${\cal O}\left(r^{1-2\nu)/2}\right)$. Both these solutions
converge, and are normalizable only if $2\nu < 1$, that is, if
$4\lambda< 1+(n-2)(10-n)$. However, in this limit, there may be a
continuum of negative energy bound states due to a negative (or
unbounded) potential. Fortunately, these solutions do not satisfy
the usual boundedness condition
\begin{equation} \int
\phi^2\,\frac{r^{n-1}}{f(r)}\,dr=1, \end{equation} because
$\phi=\Phi\,r^{-(n-1)/2}\sim r^{-n}$ diverges at the origin. Thus,
for the $k=+1$ case, a zero mass background is stable under tensor
perturbations~\cite{Gibbons02a}, and this is true also for the
$\alpha>0$ solution, in satisfying Eq.~(\ref{newconstr}). In the
latter case, the leading behavior of the solution as $x\to 0$ is
given by ${\cal O}\left(x^{-(2\beta+1+2\nu-2\beta-2)/4}\right)
\sim {\cal O}\left(r^{(1-2\nu)/2}\right)$. However, for the
positive root in Eq.~(\ref{k=1bg}), the gravitational potential is
unbounded when $n\leqslant 7$ and is unstable as in the
$\Lambda=0$ case~\cite{Deser85a}.

\subsection{Potential for extremal background}

For $\alpha>0$, the $\mu_{extr}<0$, $k=-1$ (extremal) background
does not have constant curvature. The perturbation equations would
involve terms like $a(l^6/r^8)$ and higher powers of $(l/r)$, with
some constant $a$, namely, $a\ll{1}$ when $\alpha\ll{l}^2$. This
complicates the process for finding an exact solution to the
Schr\"odinger equation. But, in the limit $\alpha\ll{l}^2$, as
well as $r^2>2\alpha$, we may get some approximate ideas about the
stability of an extremal background by inspecting the
gravitational potential.

\begin{figure}[ht]
\begin{center}
\epsfig{figure=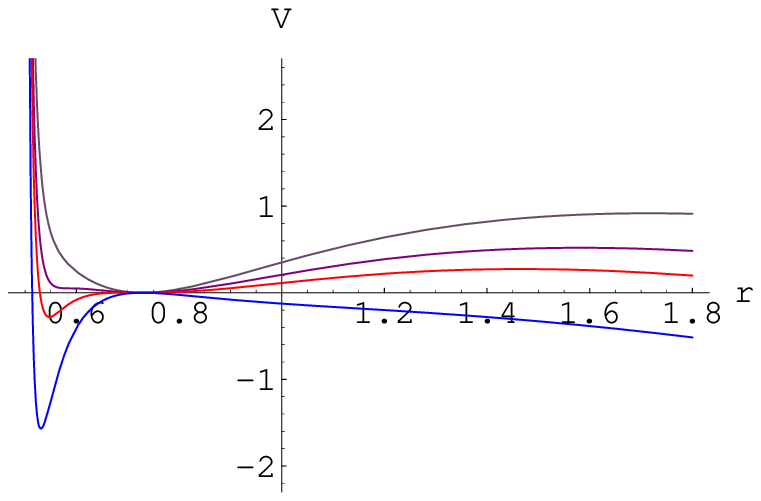,height=3.5cm,width=6.5cm}
\end{center}
\begin{center}
\epsfig{figure=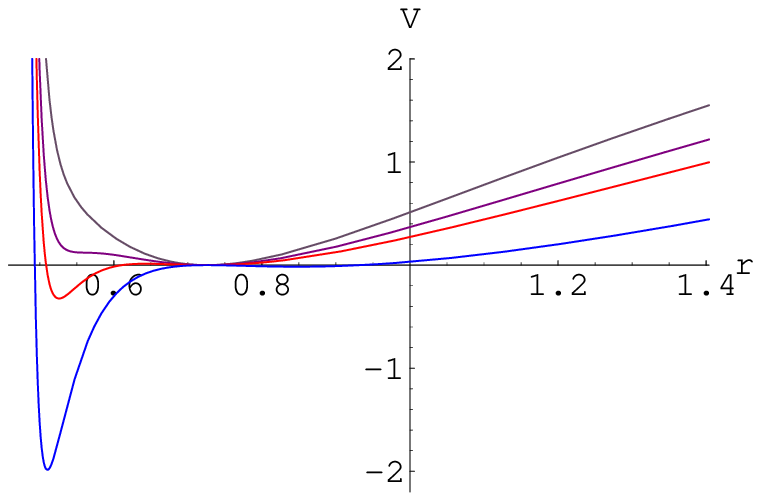,height=3.5cm,width=6.5cm}
\end{center}
\caption{The potential vs horizon radius, with an extremal state
as the ground state. The values are fixed at $n=4$, $l^2=1$,
$\lambda=-1,\,-1.6,\,-2,\,-3$ (from top to bottom); and
$\alpha=1/17$ (upper plot) and $\alpha=1/20$ (lower plot). }
\label{figure12}
\end{figure}

For $k=-1$, the gravitational potential, when $\alpha/l^2$ or/and
$r^2>2\alpha$, may be given by
 \begin{eqnarray} \label{potential5}
  V(r) &\simeq &
\Bigg[\frac{4\lambda+13}{4r^2}-\frac{17}{8\alpha}
+\sqrt{\frac{(l^2-4\alpha)\,r^4}{(r^4-\alpha l^2)\,l^2}}\nn
\\&\times& \left(\frac{17}{8\alpha}
-\frac{13\,l^4}{8(l^2-\alpha)r^4}\right)+\frac{4(2l^2-19\alpha)}
{l^2(l^2-2\alpha)}\Bigg] f(r), \nn \\
\end{eqnarray} where one reads $f(r)$ from
Eq.~(\ref{GBExtremal}). For a small coupling $\alpha\ll{l}^2$, the
gravitational potential~(\ref{potential5}) can be bounded from
below for all eigenvalues satisfying
$\lambda>\lambda_{crit}=-\,13/4$. The coupling $\alpha/l^2$ is an
equally important parameter for the dynamical stability of the
background. As the plots in Fig.~\ref{figure12} show, when
$\alpha/l^2\leqslant 1/18$ holds, the potential is bounded from
below and is positive for large $r$, but it can be negative and
unbounded for $\alpha/l^2> 1/18$. It is quite interesting that the
above constraint for $\alpha/l^2$, which we actually derived for
$k=+1$, is also effective in the $k=-1$ case.

It is possible that the form of the gravitational potential would
be modified at small scales, namely, $r^2<l^2/2$, due to terms
like $l^6/r^8$, and also due to additional higher derivative
curvature corrections, like $R^4$ terms. However, any corrections
to the potential coming from them would contribute only in the
order of $1/r^8$, so they will not destabilize the potential for
large $r$, as well as for small $r$, but in the limit
$\alpha\ll{l}^2$ and $r^2>2\alpha$.

For $k=-1$, the base manifold has negative curvature, so a
negative potential does not imply instability of the background,
if it can be bounded from below. We can be a little more precise
here. The Breitenloher-Freedman bound $\lambda>-\,n^2/4$, which
may be required for the positivity of energy (or unitarity) in
AdS$_{n+1}$ spacetime, is $\lambda>-\,4$, when $n+1=5$. In our
case, we must satisfy $\lambda> -\,13/4$, such that the potential
is bounded from below and positive for large $r$, which gives a
stronger bound for the stability of an extremal background than
the Breitenloher-Freedman bound.

A bounded and positive potential implies that there are no
unstable modes. For a small but non zero GB coupling, hyperbolic
black holes whose ground state is the extremal metric with a
negative mass may be stable under tensor perturbations. What we
have shown here is that the extremal black holes are also local
minima of the energy under small metric perturbations.

\section{Discussion and Conclusion}

In this paper, we have presented three important ideas together:
(i) the choice of a ground state for AdS black hole spacetimes
with flat, spherical, and hyperbolic horizons, (ii) the
thermodynamic stability of hyperbolic AdS black holes, and (iii)
the gravitational (or dynamical) stability of AdS black hole
spacetime of dimension $D>4$ under linear (metric) perturbations.

Having clarified which backgrounds should be used for AdS black
hole spacetimes with flat, spherical, or hyperbolic horizons, we
computed the Gauss-Bonnet curvature corrections to the AdS black
hole thermodynamics, using the standard regularization method of
Refs.~\cite{Hawking83a,Witten98a}, which follow by the subtraction
of divergences from a reference state to which the black hole
solution is asymptotically matched. The extremal black holes are
found to be local minima of the energy for anti--de Sitter black
hole spacetimes with hyperbolic event horizons.

AdS$_5$ hyperbolic black holes present some interesting features,
such as that the free energy and the entropy are dependent on the
Gauss-Bonnet coupling $\alpha$ but the total energy is not. This
equally explores the possibility that the entropy and free energy
of hyperbolic AdS black holes scale with the coupling of
$\alpha^\p$ corrections when going from strong to weak coupling
limits, as in~\cite{Emparan99b,Klemm99a}.

We have shown that, for thermodynamic stability of hyperbolic
black holes, the specific heat and the extremal entropy have to be
positive on the background. By considering free energy curves, the
corresponding thermal phase diagrams are obtained for $n=4$ and
$n=6$ in the limits of small and large Gauss-Bonnet couplings. Our
results appear to suggest that the GB type corrections to the
black hole thermodynamics do not give rise to a Hawking-Page
transition as a function of temperature for flat and hyperbolic
event horizons.

We have shown that a stable branch of small spherical black holes
will occur when the coupling $\alpha$ is small. More specifically,
in the AdS$_5$ case, a first order thermal phase transition may be
observed for a small GB coupling, namely, $\alpha<0.0278\,l^2$. In
addition, the Hawking-Page transition temperature decreases when
the coupling of $\alpha^\p$ corrections is increased. This
behavior is seen in all spacetime dimensions $D\geqslant 5$.

We have also explored the gravitational (or dynamical) stability
of higher dimensional AdS black hole spacetimes, with and without
a Gauss-Bonnet term, against metric perturbations. Our result
suggests that a base manifold ${\cal M}$ with a negative constant
curvature can be unstable under tensor perturbations if the
background is a massless topological black hole. For solutions of
the Einstein-Gauss-Bonnet theory, one may have potentials which
are positive and bounded from below, when the coupling constant
$\alpha^\prime$ is small, that is, $\alpha\ll{l}^2$.

\section*{ACKNOWLEDGMENTS}
I wish to thank Danny Birmingham, Rong-Gen Cai, Chiang-Mei Chen,
Roberto Emparan, Sean Hartnoll, Pei-Ming Ho, Miao Li, Shin'ichi
Nojiri, Sergi Odintsov, Sumati Surya and John Wang for numerous
helpful conversations and insightful comments. I would also like
to acknowledge the warm hospitality of the CERN theory group where
a part of the work was done. This work was partially supported by
the NSC and the center for Theoretical Physics at NTU, Taiwan.

\section*{Appendix: Linearized Curvature Terms for $R^2$ Gravity}
\renewcommand{\theequation}{A\arabic{equation}}
\setcounter{equation}{0}

Under a linear perturbation \begin{equation} \bar{g}_{ab}\to
g_{ab}= \bar{g}_{ab}+h_{ab}\,, \quad g^{ab}=\bar{g}^{ab}-h^{ab},
\end{equation} with $\vert
h^a\,_b\vert <<1 $, the curvatures transform as~\cite{IPN01d}
\begin{eqnarray}
R_{abc}\,^d &\to & R_{abc}\,^d
 -\bar{\nabla}_{[a}\bar{\nabla}_{|c|} h_{\,b]}\,^d
 +\bar{\nabla}_{[a}\bar{\nabla}^d
h_{b]\,c}, \nn \\
&{}&+\,\half\left(\bar{R}_{abe}\,^d h_c\,^e- \bar{R}_{abc}\,^e
h_e\,^d\right),\label{curvature}
\nn \\
R_{ab}&\to & R_{ab}+\frac{1}{2}\,\left(\Delta_L\right) h_{ab}
-\frac{1}{2}\,\bar{\nabla}_a\bar{\nabla}_b h
+\bar{\nabla}_{(a}\bar{\nabla}^c
h_{b)\,c},\label{Riccitensor} \nn \\
R &\to & R -\bar{R}_{ab} h^{ab}- \bar{\nabla}^2 h  +
\bar{\nabla}_a\bar{\nabla}_b h^{ab},\label{Ricciscalar}
\end{eqnarray} where $h=h_p^p$. The Lichnerowicz operator
$\Delta_L$ acting on a symmetric second rank tensor $h_{ab}$ reads
as \begin{equation} \left(\Delta_L
h\right)_{ab}=-\bar{\nabla}^c\bar{\nabla}_c h_{ab}-2\bar{R}_{a c b
d} h^{c d}+ 2\bar{R}_{(a}\,^c \,h_{b)\,c} \,.
\end{equation}
Diffeomorphism under $h_{ab}\to
h_{ab}+\bar{\nabla}_a\xi_b+\bar{\nabla}_b\xi_a$ implies a gauge
invariance of the linearized theory. One of the physical gauges is
the transverse (or harmonic) gauge
\begin{equation} \bar{\nabla}^a \hat{h}_{ab}\equiv
\bar{\nabla}^a\left(h_{ab}-\frac{1}{2}\,g_{ab}\,h_c^c\right)=0\,.
\end{equation} The Lichnerowicz operator is then compatible
with the transverse trace-free condition $h_c^c=\bar{\nabla} ^a
h_{ab}=0$. This gauge does not eliminate all of the gauge freedom,
but does simplify the perturbation equations \begin{equation}
\delta
R_{ab}=\frac{1}{2}\,\left(\Delta_L\right)_{ab}.\end{equation}

\subsection*{1. Background solutions}

The Einstein field equations modified by a Gauss-Bonnet term and a
cosmological constant $\Lambda$ are
\begin{equation}
R_{ab}-\frac{2\Lambda}{n-1}\, g_{ab}=16 \pi G_{n+1}\,\alpha^\p
\left(\frac{1}{n-1}\,H g_{ab} -2H_{ab}\right)\,\label{A1}
\end{equation} where $H_{ab}=R
R_{ab}-2R_{acbd}R^{cd}+R_{acde}R_b\,^{cde}-2R_{ac} R_b^c$ and
$H=H_a^a$. For $\alpha^\p>0$, the metric solution is
\begin{equation} \label{muneq0}
f(r)=k+\frac{r^2}{2\alpha}\left(1\mp \sqrt{1+
\frac{8\alpha\Lambda}{n(n-1)}+\frac{4\alpha\mu}{r^n}}\right),
\end{equation} where $\alpha=16\pi G\, (n-2)(n-3)\alpha^\prime$. For
$k=+1$, the background (AdS vacuum) is given by the setting
$\mu=0$. Then Eq.~(\ref{muneq0}) takes the form
\begin{equation} f(r)=1+C\,r^2\,, \quad C\equiv \frac{1\mp
\sqrt{1+ 8\alpha\Lambda/{n(n-1)}}}{2\alpha}\,,
\end{equation} where $C$ has the dimension
$(\mbox{length})^2$. This corresponds to a maximally symmetric
space, namely,
\begin{equation} \label{symmetricGB}
R_{abcd}=-\,C\,\left(g_{ac}g_{bd}-g_{ad}g_{bd}\right).
\end{equation}
Then the cosmological constant is fixed as
\begin{equation} \Lambda=-\,\frac{n(n-1)C}{2}\,\left(1-\alpha
C\right).\end{equation} Finally, for a spherically symmetric
solution, one has
\begin{eqnarray} f(r)&=& 1+\frac{r^2}{2\alpha}\left[1\mp
\sqrt{1-4\alpha C(1-\alpha C)}\,\right]\nn \\
&=&1+C\,r^2\,,\quad 1+\frac{r^2}{\alpha}-C r^2 \,. \end{eqnarray}
By replacing $C$ with $1/{\ell}^2$, we find $f(r)=1+r^2/{\ell}^2$.
Thus, for a spherically symmetric solution, the AdS vacuum of the
Einstein theory is also the vacuum of the Einstein-Gauss-Bonnet
theory, although the black hole solutions (which are given by
$\mu\neq 0$) with $\alpha>0$ are very different from the solutions
with $\alpha=0$.

\subsection*{2. Linearized equations}

In the background of Eq.~(\ref{symmetricGB}), the variations of
the curvature terms are given by \begin{eqnarray}
\delta\left(RR_{ab}\right)&=&
-\,n(n+1)\,C\,\delta R_{ab}-n C\,\delta R,\nn \\
\delta\left(R_{acde}R_b\,^{cde}\right)&=& -\,4 C\,\delta
R_{ab}+2C^2\left(h_{ab}-\bar{g}_{ab}\,h\right),\nn \\
\delta\left(R_a^c\,R_{bc}\right)&=&-\,2n C\,\delta R_{ab}-n^2
C^2\,h_{ab},\nn
\\
\delta\left(R_{acbd}R^{cd}\right)&=&
\frac{n C^2}{2}\left((n+3)h_{ab}-\bar{g}_{ab} h\right)\nn \\
&{}& -\,(n-1)C\,\delta R_{ab}-\bar{g}_{ab}\,C\,\delta R,
\end{eqnarray} and \begin{eqnarray} &{}&
\delta\Bigg(g_{ab}\Big(R^2-4R_{ab}R^{ab}+R_{abcd}R^{abcd}\Big)\Bigg)\nn
\\ &{}& ~~~~~~~=\, n(n-2)(n-1)(n+1)C^2\,h_{ab}\nn \\
&{}& ~~~~~~~~~~~~- \, 2(n-1)(n-2)\,C\,\delta R \,.\end{eqnarray}
Therefore, the field equations~(\ref{A1}) reduce to the form
\begin{eqnarray} \label{finallinear} \big(1-2C \alpha \big)\delta
R_{ab}&=& -\,n C(1-C\alpha)\,h_{ab}+
\frac{2(n^2+1)\alpha}{n-3}\,C^2\,h_{ab}\nn \\
&{}& -\, \frac{2\alpha\, C}{n-3}\,\Big(\delta R
+C\,\bar{g}_{ab}\,h\Big)\,.\end{eqnarray}

For an extremal metric background, however,
Eq.~(\ref{finallinear}) may be used only for large $r$. For a
small coupling $\alpha/\ell^2$, under the rescaling $l^2\to
\ell^2/(1-\alpha/\ell^2)$ in five dimensions, the extremal
solution is \begin{equation} f(r)=
-1+\frac{r^2}{2\alpha}-\frac{{\ell}^2-2\alpha}{2\alpha}
\sqrt{\left(\frac{r^4}{{\ell}^4}-\frac{\alpha}
{{\ell}^2-\alpha}\right)}\,.
\end{equation}
Therefore, in the limit $\alpha\to 0$, one has
 \begin{equation}
f(r)\approx -1+\frac{r^2}{\ell^2}+\frac{\ell^2}{4
r^2}=\frac{r^2}{\ell^2}\left(1-\frac{\ell^2}{2r^2}\right)^2.
\end{equation}
The solution with $\alpha>0$ is only asymptotically locally AdS;
viz., when $\alpha={\ell}^2/5$, the curvatures are
\begin{eqnarray} R_0^0&=& -\,\frac{4}{{\ell}^2}-
\frac{9}{64}\,\frac{{\ell}^6}{r^8}+ {\cal
O}\left(\frac{1}{r^{12}}\right)=R_1^1,\nn\\
R_{i}^{j}&=&
\left[-\,\frac{4}{{\ell}^2}+\frac{3}{64}\,\frac{{\ell}^6}{r^8}+
{\cal O}\left(\frac{1}{r^{12}}\right)\right]\delta_{i}^{j}\,.
\end{eqnarray} Although the subleading terms are much
suppressed for $r>{\ell}$, for the extremal background there are
extra contributions on the right hand side of
Eq.~(\ref{finallinear}).

\end{document}